\documentclass[12pt]{article}

\usepackage[active]{srcltx}
\usepackage{latexsym}
\usepackage{amsmath}
\usepackage{amsfonts}
\usepackage{mathrsfs}
\usepackage{dsfont}
\usepackage{verbatim}
\usepackage{xcolor}
 \csname
@addtoreset\endcsname{equation}{section}

\usepackage{lipsum}       
\usepackage{xargs}                   
\usepackage[normalem]{ulem}

\usepackage[colorinlistoftodos,prependcaption,textsize=tiny]{todonotes}
\newcommandx{\unsure}[2][1=]{\todo[linecolor=red,backgroundcolor=red!25,bordercolor=red,#1]{#2}}
\newcommandx{\change}[2][1=]{\todo[linecolor=blue,backgroundcolor=blue!25,bordercolor=blue,#1]{#2}}
\newcommandx{\Sinfo}[1]{\todo[backgroundcolor=red!25,bordercolor=red,noline]{S:#1}}
\newcommandx{\Winfo}[1]{\todo[backgroundcolor=blue!25,bordercolor=blue,noline]{W:#1}}

\def\p{\partial}

\def\a{\alpha}

\def\s{\sigma}

\def\be{\begin{equation}}
\def\ee{\end{equation}}
\def\bea{\begin{eqnarray}}
\def\eea{\end{eqnarray}}
\def\ba{\begin{array}}
\def\ea{\end{array}}

\def\tr{\text{tr}}

\def\12{\frac{1}{2}}

\def\gmt{\overset{\scriptscriptstyle{(-2)}}{g}}
\def\gz{\overset{\scriptscriptstyle{(0)}}{g}}

\def\gt{\overset{\scriptscriptstyle{(2)}}{g}}

\def\gf{\overset{\scriptscriptstyle{(4)}}{g}}
\def\gs{\overset{\scriptscriptstyle{(6)}}{g}}
\def\gn{\overset{\scriptscriptstyle{(n)}}{g}}
\def\gtn{\overset{\scriptscriptstyle{(2n)}}{g}}

\def\ez{\overset{\scriptscriptstyle{(0)}}{e}}
\def\eone{\overset{\scriptscriptstyle{(1)}}{e}}
\def\eminusone{\overset{\scriptscriptstyle{(-1)}}{e}}
\def\etwo{\overset{\scriptscriptstyle{(2)}}{e}}
\def\ethree{\overset{\scriptscriptstyle{(3)}}{e}}
\def\en{\overset{\scriptscriptstyle{(n)}}{e}}
\def\eNm1{\overset{\scriptscriptstyle{(N-1)}}{e}}

\def\psizero{\overset{\scriptscriptstyle{(0)}}{\psi}}

\def\psitwo{\overset{\scriptscriptstyle{(2)}}{\psi}}

\def\psifour{\overset{\scriptscriptstyle{(4)}}{\psi}}

\def\psisix{\overset{\scriptscriptstyle{(6)}}{\psi}}

\def\psieight{\overset{\scriptscriptstyle{(8)}}{\psi}}

\def\psitwon{\overset{\scriptscriptstyle{(2n)}}{\psi}}

\def\nablaz{\overset{\scriptscriptstyle{(0)}}{\nabla}}

\def\Rz{\overset{\scriptscriptstyle{(0)}}{R}}

\begin{document}

\begin{flushright}
\end{flushright}

\vspace{25pt}

\begin{center}


{\LARGE \bf Some aspects of holographic $\mathcal{W}$-gravity}


\vspace{25pt}
{\large Wei ~Li$^{a}$ and Stefan~Theisen$^{b}$}

\vspace{10pt}
{\sl\small
$^a$Centre for Particle Theory \& Department of Mathematical Sciences\\
Durham University, South Road, Durham DH1 3LE, UK\\
{\tt wei.li2@durham.ac.uk} 
}

\vspace{10pt}
{\sl\small
$^b$Max-Planck-Institut f{\"u}r Gravitationsphysik, Albert-Einstein-Institut\\
Am M{\"u}hlenberg 1, D-14476 Golm,\ GERMANY\\
{\tt stefan.theisen@aei.mpg.de} 

}

\vspace{70pt} {\sc\large Abstract}\end{center}
We use the Chern-Simons formulation of higher spin theories in 
three dimensions to study aspects of holographic ${\cal W}$-gravity. 
Concepts which were useful in studies of pure bulk gravity theories, 
such as the Fefferman-Graham gauge and the residual gauge transformations, 
which induce Weyl transformations in the boundary theory and their higher spin 
generalizations, are reformulated in the Chern-Simons language. 
Flat connections that correspond to 
conformal and lightcone gauges in the boundary theory are considered.  
 
\vspace{60pt}
\begin{center}
\today
\end{center}

\newpage


\tableofcontents

\section{Introduction}

Holography is a well-established powerful tool for detailed studies of conformal 
field theories. In general dimension $d$ the CFT is dual to a gravity theory in the 
$(d+1)$ dimensional bulk, possibly coupled to other bulk fields, whose boundary 
values are sources of 
certain operators in the CFT. Local symmetries in the bulk are in one-to-one 
correspondence with global symmetries on the boundary, where they can be gauged 
by coupling the theory to sources of conserved currents.   

The coupling of the CFT to an external metric leads to a diffeomorphism invariant theory, 
which, in addition, possesses classical Weyl symmetry, i.e. invariance under 
rescaling of the metric and possibly of the fields of the CFT. In two dimensions,
to which we restrict the following discussion, the 
three local symmetry parameters are sufficient to gauge away the external metric. 
In the quantum theory, the symmetries of the classical theory cannot all be 
maintained simultaneously, leading to an anomaly. It manifests itself in anomalous Ward
identities or in a non-invariance of the effective action, a functional of the 
external metic which is obtained 
by integrating out the quantum fields and generates
correlation functions of the energy-momentum tensor.

Which of the symmetries one wants to maintain dictates the choice of the counter-terms. 
 Opting for diffeomorphism
invariance (or equivalently conservation of the energy-momentum tensor) 
leads to the non-local 
Polyakov action \cite{Polyakov:1987zb}, 
whose only dependence on the specific CFT is through
an overall factor proportional to the central charge, which parametrizes the 
anomaly. In this context, the anomaly manifests itself in 
the non-invariance 
of the Polyakov action under Weyl rescaling of the metric, causing a  
non-vanishing vacuum expectation value of the trace of the energy-momentum 
tensor in an external metric background, or a non-vanishing trace of the 
two-point function of the energy-momentum tensor in flat space.   
The anomaly is also well known as the quantum-mechanically induced  
central extension of the infinite-dimensional conformal algebra (the 
symmetry algebra of classical CFT's) to the Virasoro algebra.  

Besides conformal symmetry, two-dimensional CFT's can also have enhanced 
symmetries, the most prominent ones being Kac-Moody symmetries with 
spin-one currents and supersymmetries with fermionic 
symmetry currents with spin 3/2.   
In this paper we are interested in CFT's 
with conserved
higher-spin currents whose symmetry algebras are known as 
$\mathcal{W}$-symmetries, which  have the Virasoro algebra as a sub-algebra.  
The simplest and earliest example is Zamolodchikov's $\mathcal{W}_3$ 
algebra \cite{Zamolodchikov:1985wn}. 
In the same way as a CFT can be coupled to an external 
metric, which sources the energy-momentum tensor of the CFT, 
a CFT with higher-spin ${\cal W}$-symmetries can be coupled to higher-spin gauge fields, which source 
conserved higher-spin currents.  
This leads to the notion of ${\cal W}$-gravity. 
At the classical 
level, the symmetries are higher-spin generalizations of diffeomorphism,
parametrized by a traceless symmetric rank $s-1$ tensor with two components,
and generalized 
Weyl transformations, parametrized 
by a symmetric rank $s-2$ tensor with $s-1$ components.  

In the classical theory, these 
symmetries are sufficient to gauge away the $s+1$ components of the spin-$s$ 
sources; but after quantization this is no longer possible. 
Choosing to preserve diffeomorphism invariance and higher-spin gauge symmetries 
results in anomalies in generalized Weyl symmetries. The anomalous
symmetry transformations, called ${\cal W}_s$-Weyl transformations, 
are parametrized by one scalar field for each spin $s$.
The corresponding anomalies, which we call ${\cal W}_s$-anomalies, manifest themselves
as trace anomalies in the two-point functions of higher-spin 
currents or as the non-invariance of the effective action under 
the (anomalous) 
${\cal W}_s$-Weyl transformations.
Two-dimensional theories 
with $\mathcal{W}$-algebras
as  symmetry algebra were intensively studied about 25 years ago, 
also in the context of string theory, 
but it is fair to say that the implications 
of the higher-spin symmetries are much less understood than those of the conformal 
symmetry. 
A good review of the early literature is \cite{Hull:1993kf}. 

Since then the AdS/CFT correspondence has equipped us with a new tool 
to study conformal field theories. 
To study two-dimensional 
conformal field theories with higher-spin $\mathcal{W}$-symmetries and their couplings to 
higher-spin sources, we need a three-dimensional bulk 
theory which has, in addition to diffeomorphism, higher-spin 
gauge symmetries. The source for the boundary spin-$s$ conserved 
current is the boundary value of the bulk gauge field of the same spin.

The AdS/CFT correspondence with higher-spin symmetry has recently been studied, 
in particular for the boundary dimensions $d=2,3$.  
For $d=2$, which is the dimension we are interested in this paper, the bulk 
gravity theory has an alternative description as 
Chern-Simons
theory. For 3D pure gravity, whose boundary metric sources components of the CFT 
energy-momentum tensor, its alternative description as 
$\textrm{SL}(2,\mathbb{R})\times \textrm{SL}(2,\mathbb{R})$ Chern-Simons theory has been
known for a long time \cite{AT,Witten}. More recently this was extended 
to include higher-spin bulk fields \cite{HR,CFPT1}, where a formulation 
as $\textrm{SL}(N,\mathbb{R})\times \textrm{SL}(N,\mathbb{R})$ Chern-Simons theory 
was proposed. 
Corresponding to the 
presence of the higher-spin bulk fields the boundary theory is a CFT with higher-spin ${\cal W}_N$-symmetry. 

A bulk spin-$s$ gauge field is realized  by a symmetric space-time tensor of 
rank $s$, which we will refer to as a metric-like field. 
An action principle for the interacting higher-spin theory in terms of metric-like fields is not known in general. In three dimensions, since we have an alternative description 
in terms of Chern-Simons theory, we have, in principle,  
all the information at our disposal. Given a 
pair of  flat connections, i.e. a solution of 
the equations of motion of the CS theory, we can construct 
the metric-like fields. Likewise, the 
higher-spin symmetries are encoded in the $\mathfrak{sl}(N)$ gauge symmetries, i.e. 
given a gauge symmetry we can construct the parameters of the 
higher-spin symmetries, which we refer to as generalized diffeomorphisms. 

However, for $N\geq 3$, it is unclear how to translate from connections 
to metric-like fields at the level of the action and the equations of motion.
Attempts to construct them order by order in the higher-spin fields 
were made e.g. in \cite{CFPT2,FK,CH}. 
One difficulty lies in 
determining the transformations of the 
metric-like fields under the generalized diffeomorphism, which at present can 
only be done order by order in the higher-spin fields. 
What is missing is 
an understanding of a generalization of Riemannian geometry which would allow us
to write down expressions which are covariant w.r.t. to all generalized 
higher-spin diffeomorphism, e.g. generalized curvatures.  

Even though the Chern-Simons formulation provides a complete 
description of the system, there are situations where a reformulation in terms 
of metric-like fields seems desirable. For instance, in the context of holography, 
the boundary conformal field theory is coupled to the boundary values of the 
metric-like fields in the bulk. 
The way we bypass this difficulty in this paper is to use pure gravity as a guiding 
principle, in the following sense. For pure gravity, both the metric and Chern-Simons 
formulations are well understood, therefore we can translate all well-known features, 
in particular those which are relevant in the context of holography,  
from the metric formulation to the connection one and then look for a natural 
generalization to higher-rank gauge groups, as appropriate for the 
description of higher-spin fields.

One of the earliest results in the AdS/CFT correspondence is the holographic 
description (in any even dimension) of the Weyl anomaly in terms of the dual 
bulk gravity theory, which plays the role of the non-local 
effective action  \cite{HS}. For a three-dimensional bulk, 
one can translate the analysis into the equivalent Chern-Simons formulation 
and interpret the bulk diffeomorphism, which induces a Weyl rescaling of the 
boundary metric, as a particular gauge transformation. In the same way as the 
Weyl anomaly can be interpreted as the non-invariance of the bulk action 
under bulk diffeomorphism due to the presence of a boundary, it can be alternatively 
interpreted as the non-invariance of the Chern-Simons action under gauge 
transformations, again due to the appearance of a boundary term.    
Once this is realized, a generalization to higher-rank gauge groups, 
i.e. to higher-spin theories, is possible, in the sense that ${\cal W}$-Weyl 
symmetries can be interpreted as particular gauge transformations and the 
non-invariance of the Chern-Simons action can be interpreted as the non-invariance of the 
non-local effective action of the boundary theory, thus representing the 
anomalies. The relevant gauge transformations turn out to be those generated by 
the diagonal Cartan subalgebra of the two $\mathfrak{sl}(N)$ factors. 

The outline of the paper is as follows. In the second chapter 
we reformulate many features of pure three-dimensional AdS-gravity 
in the language of SL$(2,\mathbb{R})$$\times$SL$(2,\mathbb{R})$ Chern-Simons theory. 
This is mostly a review 
of well-known facts and follows to a large extent \cite{Banados:2002ey},  
in particular in translating the Fefferman-Graham gauge for the metric 
to the connection formulation of the theory.  
We then consider different gauge choices 
for the connection which correspond to different boundary metrics for the 
dual CFT. 
In the third chapter we extend the analysis to higher-rank Chern-Simons theories. 
We make an attempt to reinterpret the gauge theory results in 
terms of the higher-spin metric-like fields. 
As the main new features (and difficulties) arise already for SL(3), 
we will restrict mostly to this case, i.e. to spin three. We  
define the Fefferman-Graham gauge and among the residual gauge transformations 
those which induce ${\cal W}$-Weyl rescaling of the boundary fields. 
We use them to compute the variation of the Chern-Simons action, which is  
interpreted as the effective action of the boundary theory. We do this in the 
same gauges which we studied in the Chapter 2.
They were also studied recently, however with different emphasis, in 
\cite{deBoer:2014sna} and \cite{Indians}, respectively.
The interpretation of our result, which also touches upon the interpretation 
of the relation between bulk and boundary fields, does not seem to be 
straightforward, though.    
In Appendix A we establish our conventions for the $\mathfrak{sl}(N)$ algebras 
and their representations. In Appendix B we collect some results for general 
$\mathfrak{sl}(N)$. 
 
\section{Spin two}\label{sec:Spin-2}

\subsection{Generalities}

Our objectives are higher spin theories. As a preparation 
we review the CS-formulation of pure three-dimensional gravity and 
state some of the relevant features in a way that suggests
a natural generalization to the higher-spin case.  

The action of three-dimensional gravity with a cosmological constant 
in the second-order formulation is 
\begin{equation}\label{EH}
S={1\over 16\pi G_N}\int_{\cal M} d^3 x \sqrt{G}\Big({\cal R}+{2\over\ell^2}\Big)
\end{equation}
where $G$ is the metric and ${\cal R}$ the Ricci scalar. 
$\ell$ is a length scale, which we will often set to one 
and $G_N$ is Newton's constant 
in three dimensions.
The action in the first-order formulation is 
\begin{equation}\label{EHfirstorder}
S={1\over 4\pi G_N}\int_{\cal M} {\rm tr}\left(e\wedge R
+{1\over 3 \ell^2}e\wedge e\wedge e\right)
\end{equation}
where $e=e_\mu^a J_a dx^\mu$ is the so$(2,1)$-valued dreibein,
$R=d\omega+\omega\wedge\omega$ the curvature 2-form,  
and $\omega=\omega_\mu^a J_a dx^\mu$ with 
$\omega^a={1\over 2}\epsilon^a{}_{bc}\omega^{bc}$
the spin connection. For further details on the notation we refer to 
Appendix \ref{sec:conventions}.
The equations of motion for $\omega$ are the vanishing of the torsion,
which allows to solve algebraically $\omega=\omega(e)$. The equations of motion 
for $e$ are then the Einstein equations for the metric.  

These formulations of three-dimensional gravity can be trivially generalized 
to arbitrary dimensions. There is, however, an alternative formulation
which does not generalize to higher dimensions,\footnote{There exist higher
odd-dimensional CS gravity theories based on 
the gauge groups $SO(d-1,2)$. Their equations of motion 
are not the Einstein equations, but equations of higher order.} 
namely in terms of 
an $\textrm{SL}(2,\mathbb{R})\times \textrm{SL}(2,\mathbb{R})$ 
Chern-Simons theory \cite{AT,Witten}.
If we denote the gauge fields of the two $\textrm{SL}(2)$ factors by 
$A$ and $\tilde A$,
respectively, the action \eqref{EHfirstorder} can be written as 
\begin{equation}\label{EH-CS}
S=S_{\textrm{CS}}[A]-S_{\textrm{CS}}[\tilde A]+S_{\textrm{bndy}}
\end{equation}
where the Chern-Simons actions are\footnote{Here and in what follows we will often 
only write expressions for $A$. 
Those for $\tilde A$, unless explicitly given, follow by putting tildes
on all fields and parameters. } 
\begin{equation} 
S_{\textrm{CS}}[A]
=\frac{k}{4\pi}\int_{\mathcal{M}}\textrm{tr}(A\wedge dA 
+\frac{2}{3}A\wedge A \wedge A)
\end{equation}
and we need to identify $k={\ell\over 4 G_N}$ and 
\begin{equation}\label{3beindef}
e={\ell\over2}\big(A-\tilde A\big)\qquad \textrm{and}\qquad
\omega={1\over 2}\big(A+\tilde A\big)
\end{equation}
The difference between the Chern-Simons action and the Einstein-Hilbert action 
is a boundary term
\begin{equation}
S_{\textrm{bndy}}=-{k\over 4\pi}\int_{{\cal M}}{\rm tr}
\left[\,d\,(A\wedge\tilde A)\right]
=-{k\over 4\pi}\int_{\p{\cal M}}{\rm tr}(A\wedge\tilde A) 
\end{equation}
The metric can be recovered from the connection via
\begin{equation}\label{metricdef}
ds^{2}=G_{\mu\nu}dx^{\mu}dx^{\nu}=2\,\textrm{tr}\big(e\otimes e\big)
\end{equation}
The equations of motion are the flatness conditions for $A$ and $\tilde A$
\begin{equation}\label{EOM}
F=dA+A\wedge A=0 
\end{equation}
They are invariant\footnote{The action is invariant only up to a 
boundary term and a topological term which, in the quantum theory, leads 
to a quantization of $k$.} under SL$(2,\mathbb{R})$ gauge 
transformations
\begin{equation}\label{UUt}
A \rightarrow U^{-1}A U+U^{-1}d U 
\end{equation}
whose infinitesimal version is $\delta A=d\lambda+[A,\lambda]$.
Here $U=\exp(\lambda)\in \rm{SL}(2,\mathbb{R})$ and 
$\lambda\in\mathfrak{sl}(2,\mathbb{R})$.

If we define 
\begin{equation}\label{zetaLambda}
\zeta\equiv {1\over2}(\lambda-\tilde\lambda)\qquad\qquad\hbox{and}\qquad\qquad
\Lambda\equiv{1\over2}(\lambda+\tilde\lambda)
\end{equation}
then the infinitesimal version of (\ref{UUt}) gives
\begin{equation}\label{diffeoLorentz}
\delta_\zeta e=d\zeta+[\omega,\zeta]\qquad \textrm{and}\qquad 
\delta_\Lambda e=[e,\Lambda]
\end{equation}
Comparing this with pure gravity \cite{Witten}
identifies $\zeta$ as the parameters of diffeomorphisms and 
$\Lambda$ as those of Lorentz transformations. 
In these expressions $\zeta$ is Lie algebra valued. 
The corresponding space-time vector $\xi$
is 
\begin{equation}\label{ximu}
\xi_\mu dx^\mu={\rm tr}\big(e\otimes\zeta) 
\end{equation}
Holographic considerations usually use the metric formulation based on
the Einstein-Hilbert action. To this end, the Fefferman-Graham (FG)
gauge for the metric, its Fefferman-Graham expansion, and the Penrose-Brown-Henneaux 
(PBH) transformations have proven very useful.
One uses diffeomorphisms to bring the metric to the 
FG form \cite{FG}
\begin{equation}\label{FGgaugeG}
~~~~~~~ds^{2}=G_{\mu \nu}dx^{\mu}dx^{\nu}=\ell^2\left(\frac{d\rho^2}{\rho^2}
+\frac{1}{\rho^2}g_{ij}(\rho,x^{i})dx^i dx^j\right)
\qquad i,j=1,\dots,d
\end{equation}
where $\rho$ is the radial coordinate and $\rho=0$ is the boundary with coordinates $x^i$. 
This gauge is particularly convenient in writing the bulk/boundary dictionary as 
there are no cross-terms $G_{\rho i}$. 
As shown in \cite{FG}, $g_{ij}(\rho, x)$ has an expansion in the vicinity 
of the boundary (FG expansion)\footnote{when the back-reaction from other 
fields can be ignored. } 
\begin{equation}\label{FGexpandG}
g_{ij}(\rho, x)=\sum^{}_{n\geq 0} \rho^{2n}\,\gtn_{ij}(x)
\end{equation}
where $\gz$ is the boundary metric. 
For even boundary dimension $d$ there are additional terms 
containing logarithms of the radial 
coordinate. In $d=2$, which is the case we are interested in, they are however 
absent. Furthermore, in $d=2$, in contrast to higher dimensions, 
the FG expansion is finite \cite{Max,SS}.\footnote{Finite FG expansions 
of higher-dimensional CS theories were 
discussed in \cite{BMT}.}
It terminates 
after the third term $\rho^2 \gf(x)$, 
which is completely fixed in terms of the lower terms as
\begin{equation}\label{g4}
\gf= {\textstyle\frac{1}{4}}\,\gt\,\gz{}^{-1}\gt
\end{equation}
The vacuum expectation value of the conserved stress-energy 
tensor $\boldsymbol{T}$ of the boundary CFT coupled to an external metric $\gz$ is \cite{dHSS}
\be\label{vevT} 
\langle \boldsymbol{T}_{ij}\rangle=k\,\left(\gt\!_{ij}-\gz\!_{ij}\tr(\gt)\right)
\equiv k\,T_{ij}=k\,( t_{ij}+q_{ij})
\ee
where $t_{ij}$ is a (non-local) functional of $\gz$ and $q_{ij}$ is 
traceless and conserved w.r.t. $\gz$. $\gz$ and $q$ are the boundary data
which specify a bulk solution. 

The FG gauge is not a complete gauge fixing. The residual diffeomorphism, 
called PBH transformations, are  
generated by those $\xi^\mu$ which satisfy ${\cal L}_\xi G_{\rho\rho}=
{\cal L}_\xi G_{\rho i}=0$, whose solution is \cite{Imbimbo:1999bj}
\begin{equation}\label{PBH}
\xi^{\rho}(\rho,x)=-\rho \, \sigma(x)\,,\qquad
\xi^{i}(\rho,x)=\partial_j\sigma(x)\int^{\rho}_0 d\rho'\rho'g^{ij}(\rho',x)+\xi^{i}(0,x)
\end{equation}
Except for the 
boundary term $\xi^{i}(0,x)$, which generates an uninteresting
boundary diffeomorphism and which will be set to zero from here on,
the PBH transformation is parametrized by 
a single function $\s(x)$ on the boundary.
A PBH transformation acts on $g_{ij}(\rho,x)$ as
\begin{equation}\label{PBHonGij}
\delta_{\xi}g_{ij}=\sigma ( 2-\rho \,\partial_\rho)\,g_{ij}
+\nabla_i\xi_j+\nabla_j\xi_i
\end{equation}
where $\nabla$ is w.r.t. $g_{ij}$ and $\xi_i\equiv g_{ij}\xi^j$. 
This implies 
\begin{equation}
\delta_{\xi}\gz_{ij} = 2\,\sigma \gz_{ij}
\end{equation}
i.e. the bulk PBH transformation induces a Weyl rescaling of the boundary metric
which integrates to $\gz_{ij} \mapsto e^{2 \sigma}\gz_{ij}$ for 
finite transformations.
It is easy to work out the PBH transformations of the higher $\gn$;
for instance, $\delta_\xi \gt_{ij}=\nablaz_i\nablaz_j\s$,
which is solved by
\begin{equation}\label{g2}
\gt_{ij}=-{1\over(d-2)}
\left(\Rz_{ij}-{1\over2(d-1)}\Rz\,\gz_{ij}\right)
\end{equation} 
The pole at $d=2$ reflects the non-locality of 
$\langle \boldsymbol{T}_{ij}\rangle$ and the fact 
that 
\begin{equation}\label{trg2}
{\rm tr}\gt=-{1\over2(d-1)}\Rz
\end{equation}
is finite reflects the locality of the 
Weyl anomaly $\langle \boldsymbol{T}^i{}_i\rangle$.
We remark that \eqref{g2} is the part of $\gt$ which can be expressed completely 
in terms of $\gz$. It does not yet contain the second set of boundary data, $q$. 
In addition, if other fields are present which allow for the construction  
of Weyl-invariant symmetric tensors, they can also contribute to 
$\gt_{ij}$ and to fix them we need to go on-shell. 

Using holography there is an easy way to compute the Weyl anomaly of the 
boundary CFT, i.e. the non-invariance of the effective action $W[g]$
under Weyl rescaling of $g$. $W[g]$, the generating function
for correlation functions of the energy-momentum tensor, is 
obtained by coupling the CFT to an external metric $g=\gz$ 
and integrating out the CFT. 
A bulk diffeomorphism leaves the dual gravity action with Lagrangian 
$\mathcal{L}$ invariant, up to a boundary term
\begin{equation}
\delta_\xi S
=\int_{\cal M}\,\p_\mu(\xi^\mu \mathcal{L})
=\int_{\partial\mathcal{M}}\,\xi^\rho \mathcal{L}
\end{equation}
If we go to FG gauge, perform a FG expansion of the integrand,
and use a PBH diffeomorphism, 
i.e. $\xi^\rho=-\rho\,\sigma$, 
the on-shell ${\cal O}(\rho^0)$ term is the anomaly \cite{ST}\footnote{We will 
often use $g$ to denote the boundary metric $\gz$ when there is no risk of confusion.}
\begin{equation}\label{deltaW-SL(2)}
\delta_\sigma W[g]=\delta_\xi S\big|_{\rho^0}=\frac{1}{2\pi}\int_{\p{\cal M}}\!\!\sqrt{g}\,\s {\cal A}
\end{equation}
Possible divergencies at $\rho=0$ are cancelled by adding local 
boundary terms to the bulk action. Applied to \eqref{EH} this gives 
${\cal A}={\ell\over 4 G_N}{\rm tr}\gt$  or
\begin{equation}\label{anomaly-SL(2)}
{\cal A}=-{c\over 12}R\qquad\hbox{with}\qquad 
c={3\ell\over 2 G_N} 
\end{equation}
where $R$ is the Ricci scalar of the boundary metric. 

We will now translate these results to the CS formulation \eqref{EH-CS}. 
Here, of course, we will have to set $d=2$. 

In the coordinates $x^\mu=(\rho,x^1,x^2)$ the connection decomposes as
\begin{equation}\label{Adecompose}
A=A_\mu dx^\mu=A_\rho d\rho+A_i dx^i~\in~\mathfrak{sl}(2,\mathbb{R})
\end{equation}
Using the invariance of the action under \eqref{UUt}
we can choose a gauge for $(A,\tilde{A})$ that best suits 
the holographic description: the analogue of the 
Fefferman-Graham gauge with $G_{\rho\rho}=1/\rho^2$ and $G_{i\rho}=0$.
It is easy to see that with
\begin{equation}\label{ArhoAtrho}
A_{\rho}=-\frac{1}{\rho}L_0 \qquad \textrm{and} \qquad \tilde{A}_{\rho}=\frac{1}{\rho}L_0
\end{equation}
\eqref{metricdef} leads to 
\begin{equation}\label{FGGrhorho}
G_{\rho\rho}=\frac{1}{\textrm{tr}[(L_0)^2]}
\textrm{tr}\Big(\frac{A_{\rho}-\tilde{A}_{\rho}}{2}\Big)^2
=\frac{1}{\rho^2}.
\end{equation} 
In \cite{CFPT1}
it was shown that this gauge choice can always be achieved with a 
group element $U$ that goes to the identity at the boundary. This condition is 
necessary if we want that any Dirichlet boundary condition 
(in \cite{CFPT1} it was $A_-=0$ at $\rho=0$) 
is preserved. 

Our gauge choice for $(A,\tilde{A})$  is therefore
\begin{equation}\label{AAtgauge}
\begin{aligned}
&A=b^{-1}\, a\,b +b^{-1}\,db\qquad\textrm{with}\qquad b\equiv e^{-(\log \rho) L_0}\\
&\tilde{A}=\tilde{b}^{-1}\,\tilde{a}\,\tilde{b}+\tilde{b}^{-1}\,d\tilde{b}\qquad 
\textrm{with}\qquad\tilde{b}\equiv b^{-1}
\end{aligned}
\end{equation}
where $(a,\tilde{a})$ are $\mathfrak{sl}(2,\mathbb{R})$-valued one-forms along the 
boundary directions: $a=a_i dx^i$ and $\tilde{a}=\tilde{a}_i dx^i$.
With this choice $F_{\rho i}=0$ leads to
\begin{equation}\label{flatrhoi}
\begin{aligned}
a=a(x)
\end{aligned}
\end{equation}
i.e. $a$ depends only on the boundary
coordinates $x^i$. The remaining flatness conditions $F_{ij}=0$ 
are simply flatness of $a$ and $\tilde{a}$:
\begin{equation}\label{flataat}
da+a\wedge a=0
\end{equation}
All information is now encoded in the connections $(a,\tilde{a})$, 
which only depends on the boundary coordinates $x^i$. A generic 
$a\in\mathfrak{sl}(2)$ can be expanded as
\begin{equation}\label{aatsl2}
\begin{aligned}
a(x)=a_i(x) dx^i& \qquad \textrm{with}\qquad 
a_i=a_i^+\,L_1+a^0_i\,L_0+\,a^-_i\,L_{-1} \\
\end{aligned} 
\end{equation}
and, using \eqref{AAtgauge}, 
\begin{equation}\label{AAtsl2}
\begin{aligned}
&A_\rho=-  {L_0\over\rho}\,,\qquad
A_i=   {1\over\rho}\,a_i^+\,L_1+a^0_i\,L_0+\rho\,a^-_i\,L_{-1} \\
&\tilde A_\rho=\phantom{-}{L_0\over\rho}\,,\qquad
\tilde A_i=\rho\,\tilde a_i^{+}\,L_{1}+\tilde a^0_i\,L_0 
+{1\over\rho}\,\tilde a^{-}_i\,L_{-1}
\end{aligned}
\end{equation}
For the dreibein $e=e_{\rho}\, d\rho+e_i \,dx^i$ defined in (\ref{3beindef})
we obtain
\begin{equation}\label{3beinsl2}
\begin{aligned}
e_\rho&=-{1\over\rho}L_0\,,  \quad e_i={1\over 2}(a_i^0-\tilde a_i^0)L_0
+\frac{1}{2}\left({1\over\rho}\,a_i^{+}-\rho\,\tilde a_i^{+}\right)L_{1}
+\frac{1}{2}\left(\rho\,a_i^{-}-{1\over\rho}\tilde a_i^{-}\right)L_{-1}
\end{aligned}
\end{equation}
With \eqref{metricdef} it is clear that the metric will not be in FG 
gauge, the culprit being the zero mode component in $e_i$, which leads to 
$G_{i\rho}\neq0$. 
To remove it we use the residual gauge freedom which preserves the 
gauge choice (\ref{ArhoAtrho}).
Making a Gauss decomposition of $U(\rho,x)$ and 
$\tilde U(\rho,x)$ (with $\a^\pm=\a^\pm(x)$, etc.)
\begin{equation}
U=e^{{1\over\rho}\a^{+}L_{1}}e^{\a L_0}e^{\rho\,\a^{-}L_{-1}}\,,
\qquad
\tilde U=e^{{1\over\rho}\tilde\a^{-}L_{-1}}e^{\tilde\a L_0}e^{\rho\,\tilde \a^{+}L_{1}}
\end{equation}
with the choice\footnote{As noted before, 
$\alpha^{+}=\tilde{\alpha}^{-}=\alpha=\tilde{\alpha}=0$ is also the condition that 
the group elements $U$ and $\tilde U$ become unity at the boundary and, furthermore, 
these transformations leave $a_i^{+}$ and 
$\tilde a_i^{-}$ invariant.} 
\begin{equation}
\begin{aligned}
&\alpha^{+}=\tilde{\alpha}^{-}=\alpha=\tilde{\alpha}=0 \,,\qquad
\begin{pmatrix} 
\alpha^{-} \\ \tilde\alpha^+ 
\end{pmatrix}
=-{1\over 2} M^{-1}
\begin{pmatrix}
a^0_1-\tilde a^0_1\\ a^0_2-\tilde a^0_2
\end{pmatrix}
\quad\quad M=\begin{pmatrix}
a^{+}_1 & a^{+}_2\\ 
\tilde{a}^{-}_1 & \tilde{a}^{-}_2
\end{pmatrix}
\end{aligned}
\end{equation}
leads to \cite{Banados:2002ey}
\begin{equation}\label{FG-sl(2)}
\tilde a_i^0=a_i^0
\end{equation}
and therefore 
removes the zero modes of the dreibein, giving
$G_{i\rho}=0$. The gauge choices \eqref{AAtgauge} and \eqref{FG-sl(2)} are 
the FG gauge condition in the CS formulation of three-dimensional gravity. 
We note that the finiteness of the FG expansion of the dreibein and the metric 
is manifest. One can show that the FG expansions of $\xi_i$ and $\sqrt{G}$ 
(but not of $\xi^i$)  
are finite as well.

With the above gauge choice, the bulk metric \eqref{metricdef}
becomes
\begin{equation}\label{Gmunu}
\begin{aligned}
&G_{\rho\rho}={1\over\rho^2}\,, \qquad G_{i\rho}=0\,,\qquad
G_{ij} =\frac{1}{\rho^2}\,\gz_{ij} + \gt_{ij}+\rho^2 \gf_{ij}
\end{aligned}
\end{equation}
with 
\begin{equation}\label{gsl2}
\begin{aligned}
\gz={\textstyle\frac{1}{2}}\,a_i^{+}\tilde a_j^{-}\, dx^i dx^j\,,\quad
\gt=-{\textstyle\frac{1}{2}}\left(a_i^{+}a_j^{-}
+\tilde a_i^{-}\tilde a_j^{+}\right)dx^i dx^j\,,\quad
\gf={\textstyle\frac{1}{2}}\,a_i^{-} \tilde a_j^{+}\, dx^i dx^j
\end{aligned}
\end{equation}
where the coefficients satisfy 
the flatness condition (\ref{flataat}). 
Using those and the FG gauge condition \eqref{FG-sl(2)},
one verifies \eqref{g4} and \eqref{trg2}.

We know from the metric formulation that the gauge fixing is not 
yet complete. Indeed, transformations parametrized by $\alpha$ and $\tilde\alpha$
have a simple effect on $a_i^\pm$ and $\tilde a_i^\pm$:
\begin{equation}\label{residual}
a_i^\pm\to e^{\pm\alpha}\,a_i^\pm\,,\qquad
\tilde a_i^\pm\to e^{\pm\tilde\alpha}\,\tilde a_i^{\pm}
\end{equation} 
If we define 
\begin{equation}
\sigma={\textstyle{1\over2}}(\alpha-\tilde\alpha)\qquad\hbox{and}\qquad
\tau={\textstyle{1\over 2}}(\alpha+\tilde\alpha)
\end{equation}
then $\sigma$ acts as a Weyl rescaling and  
$\tau$ as a Lorentz transformation of the boundary zweibein. In particular
\begin{equation}
\gz_{ij}\to e^{2\,\sigma}\,\gz_{ij}
\end{equation} 
Of course, the transformation (\ref{residual}) reintroduces $G_{i\rho}$ components, but from the 
previous discussion we know that we can transform them away without affecting 
the boundary zweibein. In the metric formulation these are the transformations 
generated by the $\xi^i$.
We therefore conclude that the transformations parametrized 
by $\sigma$ are the PBH transformations of the metric formulation. 
The remaining two parameters $\alpha^-$ and $\tilde\alpha^{-}$ parametrize
boundary diffeomorphisms.

We now discuss the holographic computation of the Weyl anomaly in the CS formulation. 
For this we apply the procedure outlined above to the action \eqref{EH-CS}.
On-shell a diffeomorphism of $A$ can be written as a gauge transformation,   
\begin{equation}
\delta_\xi A={\cal L}_\xi A=d\lambda+[A,\lambda]
=\delta_\lambda A 
\qquad\textrm{with}\qquad \lambda=\imath_\xi A
\end{equation}
and likewise $\delta_\xi\tilde A=\delta_{\tilde\lambda}\tilde A$ 
with $\tilde\lambda=\imath_\xi\tilde A$. 
Under such a transformation with $\xi^\mu$ being the PBH 
diffeomorphism, the action changes as\footnote{There is a divergent (${1\over\rho^2}$)
term proportional to $\sigma\sqrt{g}$.}  
\begin{equation}
\begin{aligned}\label{deltaS}
\delta_\s W
&=\phantom{-}{k\over2\pi}\int_{\p{\cal M}}{\rm tr}\big[\s L_0(dA+d\tilde A)\big]\\
&=\phantom{-}{k\over 4\pi}\int_{\p{\cal M}}\sigma\left(\p_i a_j^0-\p_j a_i^0\right)
dx^i\wedge dx^j
\end{aligned}
\end{equation}
Bulk and boundary term in \eqref{EH-CS} give equal
contributions. This was also observed in 
\cite{RS}.\footnote{Other boundary 
terms such as those used in \cite{Banados:2002ey} lead to the same anomaly.}
Using \eqref{gsl2} and the on-shell and FG gauge conditions, one finds
\begin{equation}\label{sl2anomaly}
\delta_\s W[g]=-{k\over 4\pi}\int_{\p{\cal M}}\!\!d^2 x\,\sqrt{g}\,\s\,R
\end{equation}
Comparing \eqref{sl2anomaly} with \eqref{deltaW-SL(2)} and 
\eqref{anomaly-SL(2)} 
we verify the known relation $c={6\,k}$. 

We note that \eqref{deltaS} is nothing but (the ${\cal O}(\rho^0)$
term of) the change of the action under a gauge transformation with 
parameter $\lambda=\sigma L_0=-\tilde\lambda$. This was expected from 
the discussion above, where we found the relation between 
PBH transformations and $\mathfrak{sl}(2)$ gauge transformations. 

So far the discussion has been completely general. We will now consider special choices 
of the boundary metric and translate them into the CS formulation.
This discussion follows largely \cite{Banados:2002ey}. 

\subsection{Conformal gauge}

The first case is $\gz_{ij}=e^{\Phi}\delta_{ij}$, i.e. the boundary metric is 
in conformal gauge. In this case, the 
non-local Polyakov action $W[g]$, which is completely fixed up  
to a multiplicative constant
$c$ (the central charge of the CFT), becomes the (local) Liouville action for 
$\Phi$ and correlation functions of the 
energy-momentum tensor are expressed in terms of the Liouville field $\Phi$.  
In this gauge $\gt$ in \eqref{g2} has a well-defined limit in $d=2$
\be
\gt_{ij}={\textstyle{1\over 2}}\p_i\p_j\Phi-{\textstyle{1\over 4}}\p_i\Phi\,\p_j\Phi
+{\textstyle{1\over8}}\delta_{ij}(\p\Phi)^2 +q_{ij}
\ee
where we have added a conserved and traceless $q_{ij}$.
Via \eqref{g4} the bulk metric is now completely fixed in terms of 
$\Phi$ and $q$.
Choosing a complex structure on the boundary\footnote{This is appropriate if the 
boundary has Euclidean signature. But we will use this terminology also for 
Minkowskian signature where the `complex' coordinates should be thought of 
as light-cone coordinates and the hermitian conjugation of the connection 
acts on the Lie-algebra generators and replaces $a_i^+$ by an independent
function $\tilde a_i^-$. In this case the complex structure 
should be interpreted as a light-cone structure.} 
 s.t. $ds^2=e^{\Phi}dz d\bar z$,
one finds for $T$ of \eqref{vevT}
\begin{equation}
t_{zz}=-{\textstyle{1\over4}}(\p\Phi)^2
+{\textstyle\frac{1}{2}}\p^2\Phi\,\qquad
t_{\bar z\bar z}=-{\textstyle{1\over 4}}(\bar\p\Phi)^2+\frac{1}{2}\bar\p^2\Phi\,\qquad
t_{z\bar z}={\textstyle\frac{1}{2}}\p\bar\p\Phi
\end{equation}
and
\be
q_{zz}=q(z)\,\qquad q_{\bar z\bar z}=\bar q(\bar z)\,,\qquad q_{z\bar z}=0
\ee
We recognize $t_{zz}$ and $t_{\bar{z}\bar{z}}$ as the (traceless) 
energy-momentum tensor 
of Liouville theory.\footnote{$t_{zz}$($t_{\bar{z}\bar{z}}$) 
is (anti)holomorphic if we impose the Liouville equation 
$\p\bar\p\Phi=\mu \exp(\Phi)$, but in the present context there is no reason 
to do so.} 

The bulk metric with $\Phi\neq 0$ can be obtained from the one with $\Phi=0$
by a finite PBH transformation. As we will now show, this can be easily 
translated to the CS-formulation of pure gravity 
(and generalized to the higher-spin case, cf. Section \ref{sec:Spin3}).
In FG gauge, the flat connections $(a,\tilde{a})$ 
that correspond to the on-shell bulk metric in conformal gauge
are \cite{Banados:2002ey}
\begin{equation}
\begin{aligned}\label{a1}
a_z&=e^{\phi}\, L_{1}-\p\tilde{\phi}\, L_{0}-e^{-\phi}\, T \, L_{-1}\,, 
\qquad a_{\bar{z}}=\bar{\p} \phi \, L_{0}
+{\textstyle\frac{1}{2}}e^{-\phi}\, R\, L_{-1}
\\
\tilde{a}_z&=-\p\tilde{\phi}\, L_{0}
+{\textstyle\frac{1}{2}}e^{-\tilde{\phi}}\, R\, L_{1}\,, 
\qquad
\tilde{a}_{\bar{z}}=e^{\tilde{\phi}}\,L_{-1}+\bar{\p} \phi \, 
L_{0}-e^{-\tilde{\phi}}\, \bar{T}\, L_{1}
\end{aligned}
\end{equation}
where $(T,\bar T)\equiv (T_{zz}, T_{\bar z\bar z})$, and 
$R\equiv R_{z\bar z}=-\p\bar\p\Phi$ is the boundary Ricci tensor.
The two fields $\phi$ and $\tilde\phi$ satisfy
\begin{equation}\label{split}
\phi +\tilde{\phi}=\Phi 
\end{equation}

Consider the pair of connections
\begin{equation}\label{freevev}
a_0=[L_1- q(z)L_{-1}] dz  \qquad \textrm{and}\qquad 
\tilde{a}_0=[L_{-1}- \tilde{q}(\bar{z})L_{1}] \, d\bar{z}
\end{equation}
It is obviously flat and in FG gauge. It corresponds 
to a flat boundary metric and vev's 
$(q(z),\bar{q}(\bar{z}))$ in the absence of the source. 
The flat connection (\ref{a1}) is related to (\ref{freevev})
via a gauge transformations with
\begin{equation}\label{sl2conformalGT}
g=e^{-{1\over2}\p\Phi L_{-1}}e^{\phi L_0} \qquad \textrm{and}\qquad 
\tilde{g}=e^{-{1\over2}\bar\p\Phi L_1}e^{-\bar\phi L_0}=(g^{-1})^\dagger
\end{equation}
Note that this gauge transformation does not depend on $(q(z),\bar{q}(\bar{z}))$. The bulk metrics derived from \eqref{a1} and \eqref{freevev} 
are related by a finite 
PBH transformation generated by \eqref{sl2conformalGT}.
The zero mode parts in $(g,\tilde{g})$ introduce the conformal mode,
while the other factors in $(g,\tilde{g})$ restore  
the FG gauge. As we have already remarked before, 
this does not modify the leading terms in the FG expansion. 

Finally, applying \eqref{deltaS} to the connections \eqref{a1} 
we obtain the conformal anomaly 
\begin{equation}\label{Weylconformalsl2}
\begin{aligned}
\delta_{\sigma} W=\frac{ k}{2 \pi}\,\int_{\p{\cal M}}\sigma\, 
\partial \bar{\partial }\Phi\,dz\wedge d\bar{z}
\end{aligned}
\end{equation}
Using that the Ricci scalar for the conformal metric is
$R=-4e^{-\Phi}\partial\bar{\partial} \Phi$, we confirm that this agrees 
with \eqref{sl2anomaly}. 
This can be integrated to $W={k\over 8\pi}\int_{\p M}\Phi\,\p\bar\p\Phi$
if $\delta_\sigma\Phi=2\sigma$, which is the Weyl rescaling of the boundary metric 
in conformal gauge.

\subsection{$\mu$ - gauge}\label{sec:sl2-mugauge}

The second gauge choice which we will discuss is constructed such that the
boundary metric is
\begin{equation}\label{metriccplx}
ds^2= |dz+\mu \, d\bar{z}|^2
\end{equation}
where the Beltrami differential $\mu\equiv \mu^{z}_{\bar{z}}(z,\bar{z})$ 
defines the complex structure.\footnote{We are 
not concerned with global issues of the boundary.}  $(\mu, \bar{\mu})$ source the  
$(T_{zz},T_{\bar{z}\bar{z}})$ components of the CFT energy-momentum tensor.
The most general metric can be written as $e^{\Phi}|dz+\mu\,d\bar z|^2$
and our gauge fixing amounts to setting the conformal factor to one. We can restore it via a PBH transformation. 

We will now construct the $\mathfrak{sl}(2)\oplus\mathfrak{sl}(2)$ connection 
$(a,\tilde a)$ from which we can construct the bulk metric in FG gauge 
with \eqref{metriccplx} as boundary metric.
Recall that the Beltrami differential $\mu$ parametrizes the
complex structure $u$ on the boundary. Demanding $du=\lambda(dz+\mu\,d\bar z)$ 
requires $u$ to satisfy the Beltrami equation
\begin{equation}\label{isothe}
(\bar{\partial}-\mu \, \partial )\,u=0
\end{equation}
The pair $(\mu, T)$ with 
\begin{equation}\label{Schwarzian}
T=-\frac{1}{2}\{u,z\}\qquad\hbox{where}\qquad
\{u,z\}=\p\left({\p^2 u\over\p u}\right)-{1\over2}\left({\p^2 u\over \p u}\right)^2
\end{equation}
defines a projective structure. 
Given $T$ and $\mu$, 
the consistency between (\ref{isothe}) and (\ref{Schwarzian}) 
requires them to satisfy the anomalous Virasoro Ward identity\footnote{We recall 
that the stress energy $\boldsymbol{T}=k T$ where $k=\frac{c}{6}$; 
hence (\ref{Schwarzian}) agrees with the usual Virasoro Ward identity.} 
\begin{equation}\label{ward2}
(\bar{\partial}-\mu\,\partial -2\,\partial \mu )T = -\frac{1}{2}\partial^3 \mu
\end{equation}
whose l.h.s. is proportional to $\bar{\partial}_{\bar u}T_{uu}$ 
and the general solution to the non-anomalous 
Ward identity is therefore $(\p u)^2 q(u)$. 

The following observation connects this to the 
$\mathfrak{sl}(2)$ CS theory \cite{Verlinde:1989ua}. 
Consider the linear system
\begin{equation}\label{LSsl2}
\nabla \Psi=0 \qquad \quad \textrm{with}\qquad \nabla 
= dz\otimes(\partial +a_z ) 
+d\bar{z}\otimes(\bar{\partial} +a_{\bar{z}})\qquad \text{and}\quad \Psi= 
\begin{pmatrix}
\tilde\psi \\
\psi
\end{pmatrix}
\end{equation}
where $a$ is an $\mathfrak{sl}(2,\mathbb{C})$ connection with
\begin{equation}\label{sl2achiral}
\begin{aligned}
a_z&\equiv L_{1}-T L_{-1} 
=
\begin{pmatrix}
~0 & -T\\
-1& ~0
\end{pmatrix}\,, \qquad
a_{\bar{z}} \equiv\mu L_{1}+\bar{\omega} L_{0} + \beta L_{-1}
=
\begin{pmatrix}
\frac{\bar{\omega}}{2} & \beta\\
-\mu& -\frac{\bar{\omega}}{2}
\end{pmatrix}
\end{aligned}
\end{equation}
The holomorphic and anti-holomorphic parts of $\nabla\Psi=0$ 
imply a second-order holomorphic equation and a first-order mixed one for $\psi$,
respectively:
\begin{equation}\label{eqpsi}
(\p^2-T)\,\psi=0 \qquad \textrm{and} \qquad (\bar{\partial}
-\mu \,\partial-\frac{1}{2}\bar{\omega})\,\psi=0
\end{equation} 
Compatibility between the holomorphic and the anti-holomorphic 
parts requires $a$ to be flat.
This in turn implies two algebraic equations 
\begin{equation}\label{flat1sl2C}
\bar{\omega}=-\,\partial\mu \qquad \textrm{and}\qquad 
\beta=-{1\over 2}\partial\bar{\omega}-\mu\, T
\end{equation}  
and one first-order ODE, which is precisely the Virasoro Ward identity \eqref{ward2}. 
It is then straightforward to show that the ratio of two linearly independent 
solutions of (\ref{eqpsi}), i.e. 
\begin{equation}
u={\psi_1\over\psi_2}
\end{equation}
is a solution of \eqref{isothe} and \eqref{Schwarzian}. 
The linear system \eqref{LSsl2} is therefore equivalent to
\eqref{isothe} and \eqref{Schwarzian} \cite{Gunning}.

To derive (\ref{isothe}) and (\ref{Schwarzian}) from the flat connection 
$a$ (\ref{sl2achiral}) we rewrite $a$ as a pure gauge $a=g^{-1}\, dg$ and make a 
Gauss decomposition of $g$:
\begin{equation}
g=e^{f_{+}L_{1}}\, e^{-f_{0}L_{0}} \, e^{f_{-}L_{-1}}
\end{equation}
The minus sign in front of $f_0$ is for later convenience. The condition for
$a_z=g^{-1}\partial g$ to be in the highest weight gauge (\ref{sl2achiral}) 
gives
\begin{equation}
f_{-}=\frac{1}{2}\partial f_{0}
\qquad \textrm{and}\qquad e^{f_0}=\partial f_{+}
\end{equation}
Then $T$ can be read off from the $L_{-1}$ direction of $a_z$:
\begin{equation}
T=\frac{1}{4}(\partial f_0)^2 - \frac{1}{2} \partial^2 f_0 =-\frac{1}{2}\{f_{+}, z\}
\end{equation}
and $\mu$ can be read off from the $L_1$ direction of $a_{\bar{z}}$:  
$\mu=\frac{\bar{\partial}f_{+}}{\partial f_{+}}$.
$T$ is the Schwarzian derivative of $f_{+}$, which satisfies the Beltrami 
equation (\ref{isothe}). ($T$ is also the negative of the energy of Liouville theory 
with $\Phi=f_{0}$.)

One is now tempted to use the connection \eqref{sl2achiral} and its anti-holomorphic
counterpart $\tilde a=-a^\dagger$ to construct the bulk metric with boundary metric 
\eqref{metriccplx}. However, this fails 
because the metric would not be 
in FG gauge. The latter is obvious as the zero modes of $a$ and $\tilde a$ 
are not coupled. This can be cured with a gauge transformation generated by 
\begin{equation}\label{CtoNCsl2}
g=e^{-\frac{1}{2}\omega_{\rm FG} L_{-1}}\qquad \textrm{and}\qquad \tilde g=\big(g^{-1}\big)^\dagger=
e^{-{1\over2}\bar\omega_{\rm FG}L_1}
\end{equation}
This leads to 
\begin{equation}\label{sl2aNC}
\begin{aligned}
a_z&=L_{1}- \omega L_{0}- T L_{-1}
\,,  \qquad \qquad~
a_{\bar{z}}=\mu L_{1} +\bar{\omega} L_{0} +\beta L_{-1}\\
\tilde{a}_z&=\bar{\mu} L_{-1} -\omega L_{0} +\bar{\beta} L_{1}\,, \qquad \qquad
\tilde{a}_{\bar{z}}
=L_{-1}+ \bar{\omega} L_{0}- \bar{T} L_{1}
\end{aligned}
\end{equation}
where we have chosen $\omega=\omega_{\rm FG}+\bar\mu\,\bar\omega_{\rm FG}$. We
have also redefined $\beta$ and dropped the subscript on $\omega_{\rm FG}$. 
We will call the gauge \eqref{sl2aNC} the $\mu$-gauge. 

The flatness condition leads to two algebraic equations for $\omega$ and $\beta$
\begin{equation}\label{omegasl2NC}
\mu \omega+\bar{\omega}=-\partial \mu 
\qquad  \textrm{and}\qquad 
\beta= -{1\over2}\big(\bar{\partial} \omega
+\partial \bar{\omega}\big) -\mu T
\end{equation}
The first, combined with its anti-holomorphic counterpart, gives 
\begin{equation}\label{muomega}
\omega=\frac{\bar{\mu}\partial \mu -\bar{\partial}\bar{\mu}}{1-|\mu|^2}
\end{equation}
The last relation following from flatness is the 
anomalous Virasoro Ward identity (\ref{ward2}) 
\begin{equation}\label{wardN2}
(\bar{\partial}-\mu \partial -2\partial \mu )(T-\mathcal{Q}_2) 
= -\frac{1}{2}\partial^3 \mu\qquad\hbox{with}\qquad
{\cal Q}_2=-{1\over4}\omega^2+{1\over2}\p\omega 
\end{equation}
The shift of $T$ by $-\mathcal{Q}_2$ simply undoes the shift of $T$ 
when going from \eqref{sl2achiral} to \eqref{sl2aNC}. The ambiguity in $T$, 
previously denoted by $q$, is a solution of the non-anomalous Ward identity.  
The Ward identity obeyed by $\bar{T}$ is the complex conjugate of (\ref{wardN2}).

From (\ref{sl2aNC}) we can compute the dreibein
and the bulk metric, for which we find
\begin{equation}\label{gexpandsl2}
\begin{aligned}
&\gz=|dz+\mu d\bar{z}|^2\,, \qquad 
\gt=\frac{1}{2}[(T dz - \beta d\bar{z})(dz+ \mu d\bar{z})+{\rm c.c.}]\,,\qquad 
\gf=|T dz - \beta d\bar{z}|^2\,.
\end{aligned}
\end{equation}
with $\beta$ as in (\ref{omegasl2NC}). 

As a check we compute, using \eqref{deltaS}, 
\begin{equation}\label{PBHsl2NC}
\delta_{\sigma} W[g]=\frac{k}{2\pi} \int_{\partial \mathcal{M}}\!\! 
\sigma \,(\partial \bar{\omega}+\bar{\partial} \omega)\,d^2 z 
\end{equation}
Using \eqref{muomega} one finds $\bar\p\omega+\p\bar\omega=-{1\over2}\sqrt{g}R[g]$,   
which shows that the Weyl anomaly is correctly reproduced. 

We can also write down the bulk metric with the boundary metric in lightcone gauge 
\begin{equation}\label{LCmetric}
ds^2=(dz+\mu\, d\bar z)d\bar z
\end{equation}
by setting $\bar{\mu}=0$. This results in the connections 
\begin{equation}\label{sl2aNCchiral}
\begin{aligned}
a_z&=L_{1}- T L_{-1}
\,,  \qquad \qquad
a_{\bar{z}}=\mu\, L_{1} +\bar{\omega} L_{0} +\beta L_{-1}\\
\noalign{\vskip.2cm}
\tilde{a}_z&=-\textstyle{\frac{1}{2}}\partial\bar{\omega} \,L_{1}\,, 
\,\,\,\, \qquad \qquad \,\tilde{a}_{\bar{z}}
=L_{-1}+ \bar{\omega} L_{0}- \bar{T} L_{1}
\end{aligned}
\end{equation}
The equation of motion obeyed by 
$(\bar{\omega}, \beta)$ reduces from (\ref{omegasl2NC}) back to (\ref{flat1sl2C}). 
Note however that the holomorphic $a$ and anti-holomorphic $\tilde{a}$ are still
coupled, so as to ensure FG gauge. 

The Ward identity satisfied by $T$ reduces to the chiral one (\ref{ward2}), 
whereas the one satisfied by $\bar{T}$ is 
\begin{equation}\label{wardN2chiral}
\begin{aligned}
\partial \, 
(\bar{T}-\bar{\mathcal{Q}}_2(\bar{\omega}))  = 0 
\end{aligned}
\end{equation}
with $\bar{\omega} =-\partial \mu$. 
The Weyl anomaly is simply 
\begin{equation}\label{PBHsl2NCchiral}
\mathcal{A}_2= \frac{c}{3} \, \partial^2 \mu =-\frac{c}{12} \,R[g]
\end{equation}

\section{Spin three and higher}
\label{sec:Spin3}
\subsection{Generalities}

The goal of the previous chapter was to establish relations between the 
metric formulation of three-dimensional gravity and its connection formulation 
as a Chern-Simons theory. 
In this chapter we will turn our attention to higher-spin theories in three 
dimensions. Here we have only limited 
knowledge of its formulation in terms of the metric and higher-spin 
fields and we thus 
have to resort to its CS formulation. Much of the discussion of 
Chapter \ref{sec:Spin-2}, which was largely 
review and reformulation of well-known results,
was presented in such a way that it can be straightforwardly 
generalized from $\mathfrak{sl}(2)$ to $\mathfrak{sl}(N)$. 
However, since explicit expression become rather cumbersome, 
we will often restrict to $\mathfrak{sl}(3)$. 

The relation between the CS formulation and the formulation in terms of metric-like
fields was studied recently, see for instance \cite{CFPT2,FK,CH}, 
but here the emphasis is on different aspects than in those references. 

The description of higher-spin theories in three dimensions as higher-rank 
Chern-Simons theories was established in \cite{HR,CFPT1,CFP}. The spectrum of 
spins depends on the embedding of the gravitational
$\textrm{SL}(2,\mathbb{R})\times \textrm{SL}(2,\mathbb{R})\hookrightarrow G\times G$.
Here we will only consider $G=\textrm{SL}(N,\mathbb{R})$ and the principal embedding 
$\textrm{SL}(2,\mathbb{R})\hookrightarrow \textrm{SL}(N,\mathbb{R})$. 

The staring point for our discussion is the
action \eqref{EH-CS} with $A\in\mathfrak{sl}(N,\mathbb{R})$. 
In order for the gravity subsector to match the Einstein-Hilbert action, we need 
\begin{equation}\label{kc}
k=\frac{\ell}{4G_N}\frac{1}{2\,\textrm{tr}[(L_0)^2]}=\frac{c}{12\,\textrm{tr}[(L_0)^2]}
\end{equation}
where $\ell$ is the AdS radius, which we will often set to one, and   
$c$ is the central charge of the boundary CFT. The group theory notation is 
explained in Appendix \ref{sec:conventions}.

The generalized dreibein and spin-connection are again given by \eqref{3beindef}, 
but now they are elements of $\mathfrak{sl}(N,\mathbb{R})$. We could also 
rewrite the action in terms of those fields (see e.g. \cite{CFPT1}), but we will 
instead use \eqref{EH-CS}, which is more systematic and elegant. 
The equations of motion are again the flatness conditions for $A$ and $\tilde A$, 
i.e. $F(A)=0$ and $F(\tilde A)=0$, and they are invariant under 
$\textrm{SL}(N)$ gauge transformations.

The parameters $\zeta$ and $\Lambda$, defined as in \eqref{zetaLambda}, 
now parametrize generalized diffeomorphisms
and Lorentz transformations. Given $\zeta$, the generators of 
diffeomorphism and spin-$3$ transformation are (cf. \eqref{ximu}) 
\begin{equation}\label{xi2and3}
\xi_\mu={\rm tr}\big(e_\mu\zeta)\qquad\hbox{and}\qquad
\xi_{\mu\nu}={\rm tr}(e_{(\mu} e_{\nu)}\zeta)
\end{equation} 
and similarly for $s \geq 4$ gauge transformations. 

For principal embedding, which is essentially unique, 
there is one spin-$s$ fields $\Psi^{(s)}$ 
for $s=2,\dots,N$, one for each Casimir invariant;  
they are totally symmetric rank-$s$ space-time tensors with 
additional constraints, e.g. double tracelessness
in the free theory.
They can all be constructed from $e$. The metric $(s=2)$ is 
\begin{equation}\label{metricdefN}
ds^{2}=G_{\mu\nu}dx^{\mu}dx^{\nu}=\frac{1}{\textrm{tr}\big[(L_{0})^2 \big]}
\textrm{tr}\big(e\otimes e\big)
\end{equation}
where the normalization has been chosen to make it independent of the normalization 
of the generators of $\mathfrak{sl}(N)$.
Similarly, the spin-$3$ field is the unique (up to a normalization) 
totally symmetric rank-$3$ tensor which can be constructed from $e$:
\begin{equation}\label{spin3def}
\Psi=\Psi_{\mu\nu\lambda}dx^{\mu}dx^{\nu}dx^{\lambda}
=\frac{2}{3}\,\textrm{tr}\big(e\otimes e \otimes e\big) 
\end{equation}
By construction the higher-spin fields 
are invariant under generalized Lorentz transformations. 
For $s > 3$ this criterion leaves some ambiguities; e.g. for 
$s=4$ field, any linear combination 
\begin{equation}\label{spin4def}
\textrm{tr}\big(e\otimes e \otimes e \otimes e\big)
+c\,\big(\textrm{tr}(e \otimes e)\big)^2
\end{equation}
is a Lorentz invariant symmetric rank-$4$ space-time tensor.

The invariance of the action under \eqref{UUt} can again be used to choose the gauge
\eqref{ArhoAtrho} which, in addition to \eqref{FGGrhorho} also
implies $\Psi^{(3)}_{\rho\rho\rho}=0$. This is necessary if we 
want to have a pure gravity limit. 
In fact, if we require $A_\rho=-\tilde A_\rho$ (i.e. symmetry between the two 
connections) and that in the pure gravity case (i.e. when we switch off all higher-spin fields) $A_\rho$ and $\tilde A_\rho$ reduce to \eqref{ArhoAtrho}, this is 
the only choice. For $s=4$ this requirement e.g. 
fixes the coefficient $c$ in \eqref{spin4def}. We can therefore impose 
\begin{equation}\label{phirho0}
\Psi^{(s)}_{\rho \dots \rho}=0
\end{equation}
as part of our FG gauge condition.
As shown in \cite{CFPT1} this gauge choice is alway possible and can be achieved by 
a group element that goes to the identity at the boundary $\rho=0$. 
Our gauge choice for $(A,\tilde{A})$  is therefore again \eqref{AAtgauge}
where $(a,\tilde{a})$ are now $\mathfrak{sl}(N,\mathbb{R})$-valued and $F_{\rho i}=0$ 
lead again to \eqref{flatrhoi} and \eqref{flataat}.
The mode expansions \eqref{aatsl2} are generalized to 
\begin{equation}\label{aatmodes}
a=\sum^{N}_{s=2}\sum^{s-1}_{m=-s+1}a_{(s)}^{ m}\, W^{(s)}_m \qquad \qquad \tilde{a}
=\sum^{N}_{s=2}\sum^{s-1}_{m=-s+1}\tilde{a}_{(s)}^{m}\, W^{(s)}_m
\end{equation}
with $a_{(s)}^{ m}$ being one-forms on the boundary. 
Note that the $\rho$-dependence is completely fixed and the residual gauge 
transformations, to be discussed next, are parametrized by functions on the boundary. 
 
The choice (\ref{AAtgauge}) has a residual gauge freedom with  
\begin{equation}\label{residualUUt}
U(\rho,x)=b^{-1}\,u(x)\,b
\qquad\textrm{and}\qquad\tilde{U}(\rho,x)=\tilde{b}^{-1}\,\tilde{u}(x)\, \tilde{b}
\end{equation}
which acts as
\begin{equation}
a\rightarrow u^{-1} a u +u^{-1}du 
\end{equation}
Make Gauss decompositions
\begin{equation}\label{uutGauss}
u=u_+ \, u_0 \, u_- \qquad \textrm{and}\qquad \tilde{u}
=\tilde{u}_{-}\, \tilde{u}_{0} \, \tilde{u}_{+}
\end{equation}
where $u_+$ is generated by all the positive modes $W_m^{(s)}$ with $m>0$, etc.
The conditions for $U$ and $\tilde U$ to go to the identity at the 
boundary are
\begin{equation}\label{UUtzerobndy}
u_+=u_0=\mathds{1}\qquad\qquad\hbox{and}\qquad\qquad \tilde u_-=\tilde u_0=\mathds{1}
\end{equation}
For $\mathfrak{sl}(2)$, $G_{\rho i}=0$  
is equivalent to the condition that $e_{i}$ has no $L_0$ component.
The generalization to $\mathfrak{sl}(N)$ is that $e_i$ has no $W^{(s)}_0$ components, 
i.e.
\begin{equation}\label{FGslN}
\textrm{tr}\big[W^{(s)}_0(a_i-\tilde{a}_i)\big]=0 \qquad \textrm{with}\quad s=2,\dots,N
\end{equation}
This can be achieved with $u=u_-$ and $\tilde u=\tilde u_+$ and leads to 
\begin{equation}\label{FGspins}
\Psi^{(s)}_{\rho\dots \rho i}=0
\end{equation}
In pure gravity we could gauge away all mixed components. This is not possible 
for the higher-spin fields.

Before we continue to compute the FG expansion of the bulk spin-s field, 
we briefly discuss what we should expect from the boundary point of view. 
A field $\Phi(x)$ with scaling dimension $\Delta$, when 
coupled to gravity, has Weyl weight $\Delta$:
\begin{equation}\label{Weylweight}
g_{ij} \rightarrow e^{2 \sigma(x)} g_{ij} \qquad \Longrightarrow \qquad 
\Phi_{\Delta}(x) \rightarrow e^{\Delta \,\sigma(x)}\Phi_{\Delta}(x)
\end{equation}
where $\sigma(x)$ is the Weyl factor. In flat space a conserved spin-$s$ 
current $W_{a_1\dots a_s}$ 
has scaling dimension, hence Weyl weight, $\Delta=2-d-s$; therefore
$W_{i_1\dots i_s}=e_{i_1}^{a_1}\cdots e_{i_s}^{a_s}W_{a_1\dots a_s}$
has Weyl weight $(2-d)$. 
Coupling $W_{i_1\dots i_s}$ to the background spin-$s$ field $\varphi^{(s)}$ via
\begin{equation}\label{coupling}
\Delta S=\int d^d x\sqrt{g}\,W_{i_1\dots i_s}\varphi^{i_1\dots i_s}
\end{equation}
and requiring Weyl invariance of (\ref{coupling}) 
fixes the Weyl weight of the source
\begin{equation}\label{phiweyl}
\varphi^{i_1\dots i_s}\to e^{-2\s}\varphi^{i_1\dots i_s}
\qquad\Longleftrightarrow\qquad
\varphi_{i_1\dots i_s}\to e^{2(s-1)\,\sigma}\,\varphi_{i_1\dots i_s}
\end{equation}
For the metric ($s=2$), which is the source for the energy-momentum 
tensor with $\Delta=d$, this is the usual Weyl rescaling. 

In the holographic description, the sources are boundary values of 
bulk fields. 
Since the Weyl rescaling of the boundary 
metric is induced by a bulk diffeomorphism
with $\xi^\rho=-\rho\,\sigma(x)$, this diffeomorphism must 
also lead to a rescaling of the boundary value of the spin-$s$ fields. 
Given their transformation under Weyl rescalings this means that 
\begin{equation}
\Psi^{(s)}_{i_1\dots i_s}(\rho,x)={\varphi_{i_1\dots i_s}(x)\over\rho^{2(s-1)}}+\dots
\end{equation}
when all components are along the boundary.
For bulk fields with mixed components 
\begin{equation}
\Psi^{(s)}_{\rho\dots\rho \,i_1\dots i_k}(\rho,x)
={\varphi_{\rho\dots\rho \,i_1\dots i_k}(x)\over\rho^{s+k-2}}+\dots
\end{equation}

For pure gravity the FG expansion of the metric can be translated to a FG expansion 
of the dreibein (or vielbein, in general). In the CS formulation it translates
into a $\rho$-expansion of the connections. In FG-gauge
\begin{equation}\label{erho}
e_{\rho}=-\frac{1}{\rho}L_0
\end{equation}
The remaining two components have the $\rho$-expansion 
\begin{equation}\label{FGexpansion-e}
e_{i}={1\over\rho}\!\sum^{N}_{n=-N+2}\!\rho^{n}\,\en\!_i 
\end{equation}
with
\begin{equation}\label{dreiexp}
\begin{aligned}
\en_i dx^i=\frac{1}{2}\sum^{N}_{s=|n-1|+1}\, \left[a_{(s)}^{ -n+1} W^{(s)}_{-n+1}
-\tilde{a}_{(s)}^{n-1} W^{(s)}_{n-1}\right]
\end{aligned}
\end{equation}
for a generic $(a,\tilde{a})$ with mode-expansion (\ref{aatmodes}).
In terms of the $\rho$-expansion of $e$, the Fefferman-Graham gauge (\ref{FGslN}) 
is that $e_i$ has no $\mathcal{O}(\rho^0)$ term.

The $\rho$-expansion of the metric-like fields can then simply be computed from their 
definitions in terms of the dreibein $e$. 
The finiteness of the FG expansion 
is an immediate consequence of the construction, e.g. for the metric
\begin{equation}\label{rhoexmetric}
G_{ij}(x,\rho)={1\over\rho^2}\sum_{n=-N+2}^N\rho^{2n}\gtn\!\!_{ij}(x)
\end{equation}
If $\gn\neq 0$ for $n<0$, the back-reaction due to the 
higher-spin fields changes the leading behavior of the metric. This is to be expected, since 
the higher-spin fields at the boundary are irrelevant perturbations of the 
boundary CFT.\footnote{This is e.g. the case for the black hole solutions constructed 
in \cite{BlackHoles,Ammon:2011nk,BCT}, 
which are, however, not in FG gauge as defined here.} 
We will later discuss special cases where the strong back-reaction 
is absent. 
The $\rho$-expansion of the bulk spin-$s$ field can also be 
easily worked out using (\ref{dreiexp}).

We now discuss residual gauge transformations which preserve FG gauge. As in the 
pure gravity case, we call them PBH transformations. In pure gravity we saw that 
they induce Weyl transformations of the boundary metric, which, in the 
CS formulation, are generated by gauge transformations along the 
Cartan direction $L_0$, accompanied by compensating gauge transformations
which vanish at the boundary and restore FG gauge.  
The generalization to $\mathfrak{sl}(N)$ are the gauge transformations along the 
Cartan directions $W^{(s)}_0$ accompanied by gauge transformations which vanish  
at the boundary and restore FG gauge. 

Infinitesimal gauge transformations which preserve \eqref{ArhoAtrho} are of the form
\begin{equation}\label{deltaAAt}
\delta A= d\lambda +[A,\lambda] \qquad \textrm{with }\qquad \lambda=b^{-1}\,\alpha(x)\,b
\end{equation}
together with the $\tilde{A}$ part; $\a$ has the mode expansion 
$\a=\sum_{s,m}\a^s_m W^{(s)}_m$.  
Demanding this to preserve  
(\ref{FGslN}) imposes the following constraints 
on the parameters in $(\alpha,\tilde{\alpha})$:
\begin{equation}\label{deltaFGslN}
\textrm{tr}[W^{(s)}_0(\delta a_i-\delta \tilde{a}_i)]=0 \qquad \textrm{with}\quad s=2,\dots,N
\end{equation}
The gauge transformation (\ref{deltaAAt}) contains the (generalized) diffeomorphism 
$\delta_{\zeta}$ and the (generalized) Lorentz transformation $\delta_{\Lambda}$ 
(\ref{diffeoLorentz}). By construction, the Lorentz transformation 
has no effect on the metric-like 
fields because all $\textrm{tr}\big[[e,\Lambda]\otimes e\dots \otimes e\big]=0$.
The effect of $\delta_{\zeta}$ on the metric-like fields can be computed 
straightforwardly 
\begin{equation}\label{bulkdiffeo}
\begin{aligned}
\delta_{\zeta}G_{\mu\nu}dx^{\mu}dx^{\nu}
&= \frac{2}{\textrm{tr}[(L_{0})^2]} \left(\textrm{tr}\left[d\zeta \otimes e\right] 
+\textrm{tr}\left[[\omega, \zeta] \otimes e\right]\right)\\
\noalign{\vskip.2cm}
\delta_{\zeta}\Psi_{\mu\nu\sigma}dx^{\mu}dx^{\nu}dx^{\sigma}
&= 2 \left(\textrm{tr}\left[d\zeta \otimes e \otimes e\right] 
+\textrm{tr}\left[[\omega, \zeta] \otimes e \otimes e\right]\right) \
\end{aligned}
\end{equation}
We emphasize that restricting the gauge transformations to lie in the gravitational 
$\mathfrak{sl}(2)\oplus\mathfrak{sl}(2)$ subalgebra does not imply that the 
corresponding transformation on the metric-like fields is an ordinary 
diffeomorphism: in (\ref{xi2and3}) the spin-$3$ transformation $\xi_{\mu\nu}$ can be non-vanishing even when $\zeta$ lies only in the $\mathfrak{sl}(2)$ spanned by $L_{0, \pm 1}$ (unless we also restrict the dreibein $e$ to the same $\mathfrak{sl}(2)$). The gauge transformation that corresponds to pure diffeomorphism is simply given by $g^{\mu\nu}e_{\mu}\textrm{tr}[e_{\nu}\zeta]$ (cf. the first equation in (\ref{xi2and3}) and eq. (3.18) of \cite{CFPT2}). 

We now separate $\zeta$ into positive, zero and negative powers of $\rho$ as 
\begin{equation}
\zeta=\zeta_{+}+\zeta_{0} +\zeta_{-}
\end{equation}
where 
\begin{equation}
\zeta_{0}=\frac{1}{2}\sum^{N}_{s=2} (\alpha^{s}_0-\tilde{\alpha}^{s}_{0}) W^{(s)}_0 
\end{equation}
Generalizing the discussion of the previous chapter we expect that $\zeta_0$ 
parametrizes Weyl and ${\cal W}$-Weyl transformations of the boundary 
fields and that $\zeta_+$ can be used to transform the connections back 
to FG gauge.\footnote{If me make a Gauss decomposition of $u=\exp(\lambda)$ and 
$\tilde{u}=\exp(\tilde\lambda)$, $\zeta_0$ receives contributions from 
$(u_0,\tilde u_0)$ and $\zeta_+$ from $u_-$ and $\tilde u_+$. 
For $N>2$, requiring FG gauge does not fix all parameters in $\zeta_+$ 
in terms of $\sigma_s$.}  
We therefore define 
\begin{equation}
\sigma_s \equiv {1\over2}\big(\alpha^{s}_0-\tilde{\alpha}^{s}_0\big)
\end{equation}
as the parameters of ${\cal W}$-Weyl transformations.

To make the discussion more concrete, we discuss in detail the case of 
$\mathfrak{sl}(3)$. With principal embedding, 
there are two metric-like fields: one spin-$2$ and one spin-$3$. 
For the following discussion it is convenient to use the 
generators $W_{\pm2},\,E_\pm\equiv E^{(1)}_{\mp},\,F_\pm\equiv E^{(2)}_{\mp}$, 
and $L_0,W_0$ defined in Appendix \ref{sec:conventions}. 

We expand $a_i$ in this basis with coefficients 
${1\over2}w_i^{(\pm2)},\, e_i^{(\pm)},\, f_i^{(\pm)},\,a_i$ and $b_i$, 
all functions of the boundary coordinates. 
Conjugation by $b$ gives $A_i$ by simply replacing $W_m\to\rho^{-m}W_m$, and 
conjugation by $\tilde{b}$ gives $\tilde{A}_i$ by replacing 
$W_m\to\rho^{m}W_m$.

The components of the metric and the spin-3 field are (cf. (\ref{metricdefN},
\ref{spin3def}))
\be
G_{\mu\nu}={1\over\textrm{tr}\left[(L_0)^2 \right]}{\rm tr}
\left(e_\mu\,e_\nu\right)\,
\qquad \textrm{and}\qquad
\Psi_{\mu\nu\lambda}=\frac{2}{3}\,{\rm tr}
\left(e_{(\mu}\,e_\nu\,e_{\lambda)}\right)
\ee
where the parenthesis in $\Psi$ denotes 
symmetrization, and the normalization
factor  $\frac{2}{3}$ for $\Psi$ is chosen for later convenience.  
As
\be
G_{\rho i}={1\over2\rho}\left(a_i-\tilde a_i\right) 
\qquad\hbox{and}\qquad
\Psi_{\rho\rho i}={1\over 2\rho^2}\left(b_i-\tilde b_i\right)
\ee
they are not yet in FG gauge.
This can be fixed via a residual gauge transformation \eqref{residualUUt}. 
For this it suffices to 
use $u=u_+$ and $\tilde u=\tilde u_-$ 
(which have no effect on the boundary zweibein). 

This being done we can easily work out the components of $G$ and $\Psi$. 
As we have already remarked, from the CS construction it is a priori 
obvious that their FG expansions are finite. We will only give the leading components 
though:
\begin{equation}\label{gij}
G_{\rho\rho}={1\over\rho^2}\,, \qquad G_{\rho i}=0\,,\qquad\hbox{and}\qquad
G_{ij}={1\over \rho^2}\sum_{n=-1}^3  \rho^{2n} \,\gtn\!_{ij} 
\end{equation}
with 
\begin{equation}
\begin{aligned}
\gmt&=-\frac{1}{4}\left[ w_i^{(+2)}\tilde w_j^{(-2)}\right] dx^i dx^j\\
\noalign{\vskip.2cm}
\gz&=-\frac{1}{4}\left[e_i^{(+)} \tilde{e}_j^{(-)}
+f_i^{(+)} \tilde{f}_j^{(-)}\right]dx^i dx^j
\end{aligned}
\end{equation}
Here and for the spin-3 field below, 
the coefficients are assumed to satisfy 
the flatness condition (\ref{flataat}).\footnote{It is easy to satisfy the flatness 
conditions for constant $a$. Given an arbitrary, but generic, say $a_1$, 
$a_2=\mu a_1+\nu (a_1^2-{\rm tr}(a_1^2)\mathds{1})$ is the generic element 
which commutes with $a_1$ and the connection $a_i\,dx^i$ is flat. 
So far the cases that have been considered in the literature are mostly with constant sources and vev's.} 
The spin-3 field with
\be
\Psi_{\rho\rho\rho}=0\,, \qquad 
\Psi_{\rho\rho i}=0\,,\qquad
\ee
and
\be\label{phiijk}
\Psi_{ijk}={1\over\rho^4}\sum^{4}_{n=0}\rho^{2n}\,\psitwon\!_{ijk}
\qquad\hbox{and}\qquad
\Psi_{\rho ij}={1\over\rho^3}\sum_{n=0}^2\rho^{2n}\psitwon\!_{\rho ij}
\ee
has the leading components
\begin{equation}
\begin{aligned}
&\psizero=\phantom{-}\frac{1}{4}\left[ w_i^{(+2)}\tilde e_j^{(-)}
\tilde{f}_k^{(-)}-\tilde{w}_i^{(-2)} e_j^{(+)}
f_k^{(+)}\right] dx^i dx^j dx^k\\
\noalign{\vskip.2cm}
&\psizero_{\rho}=\phantom{-}\frac{1}{2}\left[e_i^{(+)}\tilde{e}_j^{(-)}
-f_i^{(+)}\tilde{f}_j^{(-)}
\right]\, dx^i dx^j\\
\end{aligned}
\end{equation}
We see that generically $\gmt\neq 0$, i.e. the asymptotic behavior 
of the bulk metric is changed 
due to the strong 
back-reaction of the spin-3 field on the metric. The residual gauge 
transformations parametrized by $\s_2$ and $\s_3$
act in a simple way on the leading terms of the FG expansion of the metric-like 
fields:\footnote{The transformation of the subleading terms and 
$\psi^{\scriptscriptstyle{(2n)}}_{\rho}$ are more involved.}
\begin{equation}\label{WWeylgeneral}
\begin{aligned}
&\gmt\!\!_{ij}\to e^{4\,\s_2}\,\gmt\!\!_{ij} 
\qquad \textrm{and}\qquad 
\psizero_{ijk}\to e^{4\,\s_2}\left(\psizero_{ijk}+
{\textstyle\frac{4}{3}}\,\gmt\!\!_{(ij}\p_{k)}\,\s_3\,  \right)
\end{aligned}
\end{equation}
The boundary spin-3 field has the same Weyl weight as the metric, 
which contradicts the expectation from the boundary analysis.\footnote{In early work on $\mathcal{W}$-gravity 
(see e.g. \cite{Hull:1993kf} and references therein) 
\begin{equation}
\delta \gz_{ij} = 2\,\sigma_2 \,
\gz_{ij} \qquad \textrm{and}  \qquad 
\delta \psizero_{ijk} =4\,\sigma_2\,
\psizero_{ijk} 
+ \frac{4}{3} \, \gz_{(ij}  \partial_{k)}\, \sigma_3 \nonumber
\end{equation} 
In this case the boundary metric has Weyl weight $2$ whereas the spin-$3$ field has Weyl 
weight $4$.} 

There are two ways to proceed. One is to redefine the radial coordinate
$\rho \rightarrow \rho^2$. The metric (\ref{gij}) is still 
asymptotically AdS, but with half the original radius. The leading 
term 
$g^{\scriptscriptstyle{(-2)}}$ 
now plays the role of the boundary metric. Comparing this to the boundary metric 
$g^{\scriptscriptstyle{(0)}}$ 
in (\ref{gsl2}) 
for pure gravity, we see that here $W_{\pm 2}$ 
serve the role of $L_{\pm1}$ there. With $[W_{2},W_{-2}]= 4L_0$, 
we conclude that instead of $\{L_{0,\pm1}\}$, the gravitational $\mathfrak{sl}(2)$ 
is now  $\{{\frac{1}{2}}W_{2},{\frac{1}{2}}L_0,-{\frac{1}{2}}W_{-2}\}$. 
This was interpreted in \cite{BlackHoles,Ammon:2011nk} as a flow from a
CFT with $\mathcal{W}_3$ symmetry, triggered 
by an irrelevant operator  (i.e. the spin-3 current coupled to the boundary 
spin-3 field),  to a CFT with $\mathcal{W}^{(2)}_3$ symmetry, whose holographic dual 
is obtained by choosing a diagonal embedding 
${\rm SL}(2)\hookrightarrow{\rm SL}(3)$, rather 
than the principal embedding.
The different embedding entails a different spectrum in the boundary CFT: 
The spin-2 field (i.e. the metric) still exists and is defined 
via (\ref{metricdefN}), but now with the dreibein $e$ only spanning the new gravitational 
$\mathfrak{sl}(2)$ \cite{Castro:2011fm}. In addition, there is one spin-1 and two 
spin-$\frac{3}{2}$ fields. One can compute the FG expansion of these fields as 
in (\ref{gij}) and (\ref{phiijk}). We will not pursue this line further in this paper. 

Instead, we will proceed by observing (see the explicit expression (\ref{gij}) for 
$g^{\scriptscriptstyle{(-2)}}$) that for well-chosen connections $(a,\tilde a)$, 
i.e. when either or both of 
$w^{(+2)}_i$ and $\tilde{w}^{(-2)}_i$ vanish, the $g^{\scriptscriptstyle{(-2)}}$
term in the FG expansion of the metric is absent and the asymptotic behavior of the 
bulk metric is preserved. We will consider both options 
later.\footnote{There is yet 
another alternative. If one uses the Hamiltonian 
formulation of the theory, in coordinates $(\rho, t,\phi)$, restricting the 
$\frac{1}{\rho^4}$ term to $A_{t}$ and $\tilde{A}_t$ would not change the 
asymptotic symmetries since they are Lagrange multipliers \cite{Bunster:2014mua}.}

We will now discuss generalized Weyl anomalies. 
Here we do not have the option to discuss them in terms of the metric-like 
fields since their bulk action is not known. We therefore have to resort to the 
Chern-Simons formulation. We recall the discussion in Chapter \ref{sec:Spin-2}
and note that \eqref{deltaS} is nothing but the ${\cal O}(\rho^0)$
term of the change of the action under a gauge transformation with 
parameter $\lambda=\sigma L_0=-\tilde\lambda$. 
This was expected from our discussion of PBH transformations as specific 
gauge transformations. In this chapter we identified the ${\cal W}$-Weyl transformations
as the diagonal gauge transformations generated by the Cartan directions 
in $\mathfrak{sl}(N)$.  
This therefore leads us to conjecture the change of the effective action, which is now a non-local functional of the metric and the higher-spin boundary fields
\begin{equation}\label{deltaShs}
\delta_{\s_{s}}W={k\over 2\pi}\int_{\p{\cal M}} \, \s_{s} \,
{\rm tr}\left[W^{(s)}_0(dA+d\tilde A)\right]
\end{equation}   
as the direct generalization of \eqref{deltaS}.
This should be expressible in terms of `generalized curvatures', which respect 
the covariance of $W$  under generalized diffeomorphisms.   
In the spin-two case this is what we did in \eqref{sl2anomaly}, 
but the higher-spin geometry, which would allow us to rewrite the r.h.s. of 
\eqref{deltaShs}, is not known. Even for $s=2$, if there is a nontrivial
higher-spin background $\psi$ to which the two-dimensional field theory with 
${\cal W}$-symmetry is coupled, the anomaly 
is not proportional to the Ricci scalar, because of the presence 
of higher-spin sources and because 
they transform as well under $\sigma_2$. 
Since we only have control over the Weyl variations but 
not over the variations of the metric and the higher-spin fields separately, 
we cannot, in general, compute the trace of the energy-momentum tensor and 
the conserved higher-spin currents.
We will come back to this
when we discuss specific gauges, which is what we will do next.  

\subsection{Conformal gauge}

We start with the conformal gauge 
where we turn on only those sources (metric and higher-spin backgrounds)
that couple to the trace of the 
energy-momentum tensor and the spin-$s$ currents.
They receive a vev only through the need to regularize, which introduces 
counter-terms that break 
Weyl and ${\cal W}$-Weyl symmetries. Since the operators to 
which the sources couple vanish, the boundary CFT is not perturbed 
by any irrelevant operator and we do not encounter the strong back-reaction that 
changes the asymptotics. 

The starting point is the generalization 
of \eqref{freevev}
\begin{equation}\label{freevevslN}
a_{0}=\big(L_{1}+ \boldsymbol{q}(z)\big)\,dz \qquad \textrm{with} \qquad 
\boldsymbol{q}(z)=\sum^{N}_{s=2}\frac{q_s(z)}{t^{(s)}_{s-1}}W^{(s)}_{-s+1}
\end{equation}
and similarly for $\tilde{a}_0$. 
$\boldsymbol{q}=0$ gives the AdS$_3$ vacuum and $q_s$
is the vev of the $(dz)^s$-component of the spin-$s$ current 
of the boundary field theory in a trivial background. The normalization 
of the $q_s$ is for later convenience. The connections $(a_0,\tilde a_0)$ are
obviously in FG gauge. 

The flat connection \eqref{freevevslN} can be related to an 
asymptotic AdS solution via gauge transformations $(g,\tilde{g})$ that 
are finite on the boundary.
The gauge transformations $g\,(\tilde{g})$ that transform  
(\ref{freevevslN}) to the conformal gauge include 
negative (positive) modes and zero modes. 
We write the group element in factorized form as
\be\label{gconf}
g=M^{-1}e^{\boldsymbol{\phi}}
\ee
where the first (second) factor is the negative (zero) mode part. 
The notation will be explained momentarily. 

To find $M$, we use the fact that two-dimensional ${\cal W}$-gravity in  
conformal gauge is related to Toda theory and 
recall the following construction, well-known from the 
theory of integrable hierarchies and from ${\cal W}$-algebras 
\cite{Fateev:1987zh}: given a generic element 
$\boldsymbol{\omega}\in\mathfrak{sl}(N)$ along the Cartan subalgebra (spanned by the 
zero modes) there exists a unique 
group element $M$ which is generated by negative modes only 
and which transforms $L_1-\boldsymbol{\omega}$ into $L_1-\boldsymbol{{\cal Q}}$, where 
$\boldsymbol{{\cal Q}}=\sum^N_{s=2} \frac{\mathcal{Q}_s}{t^{(s)}} W^{(s)}_{-s+1}$ 
is entirely along the highest weight directions\footnote{In the basis we are using,
$\vec\omega$ is a diagonal matrix, ${\cal Q}$ is an upper triangular matrix with 
zeros in the diagonal, and $L_1$ (the lowest weight generator of the principally 
embedded $\mathfrak{sl}(2)\mapsto\mathfrak{sl}(N)$) has entries only 
in the first lower off-diagonal; cf. Appendix \ref{sec:conventions}.} 
of $\mathfrak{sl}(N)$, i.e. 
\begin{equation}\label{Miura}
M^{-1}(L_1 -\boldsymbol{\omega})M+M^{-1}\partial M =L_1 -\boldsymbol{\mathcal{Q}}
\end{equation}
The transformation \eqref{Miura} is known as Miura transformation. 
If we take $\boldsymbol{\omega}$ to be $\partial\boldsymbol{\Phi}$, where 
$\boldsymbol{\Phi}$ is a vector of $(N-1)$ scalar fields which we refer to 
as Toda fields, then the Miura matrix and the $\mathcal{Q}_s$ are functions of 
$\partial{\boldsymbol{\Phi}}$
\begin{equation}
M=M(\p{\boldsymbol{\Phi}}) \qquad \textrm{and} \qquad \mathcal{Q}_s
=\mathcal{Q}_s(\p\boldsymbol{\Phi})
\end{equation}
and $\mathcal{Q}_s$ are the conserved charges of Toda 
theory.\footnote{They are conserved on-shell in Toda theory.}
It is known that they 
generate a classical ${\cal W}_N$ algebra (in the same way as 
$\mathcal{Q}_2$ generates the Virasoro algebra). 
Applying the inverse Miura transformation to $a_0$ and using \eqref{Miura} gives
\begin{equation}
\begin{aligned}
&a_z \rightarrow M\big(L_1+\boldsymbol{q}\big)M^{-1} -\partial M M^{-1}
=L_1-\partial \boldsymbol{\Phi} 
+M \big(\boldsymbol{q}+\mathcal{Q}(\boldsymbol{\Phi})\big)M^{-1}\\
\noalign{\vskip.2cm}
&a_{\bar z} \rightarrow 
-\bar{\partial} M M^{-1}
\end{aligned}
\end{equation}
For the zero-mode part we split (non-uniquely, cf. \eqref{split})
\be\label{Phiphi}
\boldsymbol{\Phi}=\boldsymbol{\phi}+\boldsymbol{\bar{\phi}}
\ee
This second gauge transformation 
generates ${\cal W}_s$-Weyl transformations and, at the same time, 
brings the connection into FG gauge \eqref{FGspins}.
The combined gauge transformation generated by \eqref{gconf} generates a finite 
PBH transformation of \eqref{freevevslN}. 

We now specify to $N=3$ and refer to Appendix \ref{sec:SL(N)} for general $N$. 
Using the shorthand $W_{m}\equiv W^{(3)}_m$, we expand
\begin{equation}
\boldsymbol{\Phi}\equiv \Phi_1\, H^{(1)}+\Phi_2\, H^{(2)}=\Phi_L\,L_0+\Phi_W\,W_0
\end{equation}
which implies
\be
\Phi_L={\textstyle{1\over2}}\big(\Phi_1+\Phi_2\big)\qquad \textrm{and }\qquad
\Phi_W={\textstyle{3\over2}}\big(\Phi_1-\Phi_2\big)
\ee
The Miura matrix $M(\p\boldsymbol{\Phi})=\exp(h_-)$ is
\begin{equation}\label{Miurasl3}
h_{-}=\textstyle{\frac{1}{2}}\, \p\Phi_L \, L_{-1}
+\textstyle{\frac{1}{3}}\, \p\Phi_W \, W_{-1}
+\left(\textstyle{\frac{1}{6}}\p\Phi_L\, \p\Phi_W 
-\textstyle{\frac{1}{12}}\p^2 \Phi_W\right)\,W_{-2}
\end{equation}
It generates the two Toda charges
\begin{equation}\label{chargeToda3}
\begin{aligned}
\mathcal{Q}_2(\p\boldsymbol{\Phi})&=-(\p\Phi_1)^2-(\p\Phi_2)^2
+\p\Phi_1\,\p\Phi_2+\p^2\Phi_1
+\p^2\Phi_2\\
\noalign{\vskip.3cm}
&=-(\p\Phi_L)^2+2\,\p^2\Phi_L-{\textstyle {1\over3}}(\p\Phi_W)^2\\
\noalign{\vskip.3cm}
\mathcal{Q}_3(\p\boldsymbol{\Phi})&=
\textstyle{\frac{1}{2}}\,\p^3\Phi_1 
-\textstyle{\frac{1}{2}}\,\p^2\Phi_1\,\p\Phi_2+(\p\Phi_1)^2\,\p\Phi_2
-\p\Phi_1\,\p^2\Phi_1
\,-\,(\Phi_1\leftrightarrow\Phi_2)\\
\noalign{\vskip.3cm}
&={\textstyle{2\over3}}(\p\Phi_L)^2\,\p\Phi_W-{\textstyle{2\over27}}(\p\Phi_W)^3
-{\textstyle{1\over3}}\p^2\Phi_L\,\p\Phi_W-\p\Phi_L\,\p^2\Phi_W
+{\textstyle{1\over3}}\p^3\Phi_W
\end{aligned}
\end{equation}
and, combined with $e^{\boldsymbol{\phi}}$ the $\mathfrak{sl}(3)$ connection
in FG gauge from which we compute the dreibein and from there the metric-like 
fields. 

Since in conformal gauge there is no strong back-reaction from the spin-three field, 
the FG expansion of the dreibein (\ref{FGexpansion-e}) starts with $\ez$ as in the 
$\mathfrak{sl}(2)$ case: 
\begin{equation}\label{confdreib}
\begin{aligned}
&\overset{\scriptscriptstyle{(0)}}{e}=-L_0 \,d\rho-\frac{1}{\sqrt{2}}\sum^{2}_{i=1} 
\,\left[e^{\alpha_i} E^{(i)}_{-} dz +h.c.\right] \,, \qquad \qquad 
\overset{\scriptscriptstyle{(1)}}{e}=0\\
&\overset{\scriptscriptstyle{(2)}}{e}=\frac{1}{\sqrt{2}}\sum^{2}_{i=1}\, 
\left[e^{-\alpha_i}
E^{(i)}_{+}dt_i +h.c.\right]\,, \qquad \qquad 
\ethree=\frac{1}{2}\,[e^{-\phi_1-\phi_2} W_{-2}\, dw+{\rm h.c.}]
\end{aligned}
\end{equation}
with $(\alpha_1,\alpha_2)=(2 \phi_1-\phi_2, 2 \phi_2 -\phi_1)$.
We then find for the FG expansion \eqref{rhoexmetric} of the metric
\begin{equation}\label{gCsl3}
\begin{aligned}
&\gz=\textstyle{\frac{1}{2}}(e^{2\Phi_1-\Phi_2}+e^{2\Phi_2-\Phi_1})\, dzd\bar{z}
=e^{\Phi_L}\cosh(\Phi_W)\, dz d\bar z \\
\noalign{\vskip.2cm}
&\gt
=-\textstyle{\frac{1}{2}}(dt_1+dt_2)\,dz+{\rm c.c.}
=(\partial\bar{\partial}\Phi_L)\,dz d\bar{z}
+\textstyle{1\over4}{T}\,dz^2
+\textstyle{1\over4}\bar{T}\,d\bar{z}^2\\
\noalign{\vskip.2cm}
&\gf=\textstyle{\frac{1}{2}}e^{\Phi_2-2\Phi_1}|dt_1|^2
+\textstyle{\frac{1}{2}}e^{\Phi_1-2\Phi_2}|dt_2|^2 
\qquad \qquad
\gs=e^{-\Phi_1-\Phi_2}|dw|^2
\end{aligned}
\end{equation}
Here we have defined
\begin{equation}
dt_i\equiv-{\textstyle{1\over4}}\,{T}dz
-{\textstyle{1\over2}}\,{\partial\bar{\partial}\Phi_i}d\bar{z} 
\qquad \textrm{and} \qquad 
dw \equiv \textstyle{1\over4}\,\tilde{W}dz-\textstyle{1\over4}\,{K}d\bar{z}
\end{equation}
and 
\begin{equation}
\begin{aligned}
&T\equiv q_2+\mathcal{Q}_2\qquad\qquad\qquad W\equiv q_3
+\mathcal{Q}_3\qquad\qquad
\end{aligned}
\end{equation}
and finally\footnote{$K$ vanishes upon using Toda equation 
$\partial \bar{\partial} \Phi_1=c\,e^{2 \Phi_1- \Phi_2}$ and 
$\partial \bar{\partial} \Phi_2=c\,e^{2 \Phi_2- \Phi_1}$.}\begin{equation}
\begin{aligned}
&\tilde{W} \equiv W-\textstyle{\frac{1}{3}}\,\partial \Phi_W \,T \\
\noalign{\vskip.2cm}
&K\equiv(\p\bar{\p}\Phi_L)(\p\Phi_W )
+{\textstyle\frac{1}{3}}\,(\p\Phi_L )(\p\bar{\p}\Phi_W)
-{\textstyle\frac{1}{3}}\,\p^2 \bar{\p}\Phi_W
\end{aligned}
\end{equation}
As in the $\mathfrak{sl}(2)$ case, $(T,W)$ are related to the vev of the 
stress energy $\boldsymbol{T}$ and spin-$3$ current $\boldsymbol{W}$ by a 
rescaling: $(T,W)=\frac{1}{k}(\boldsymbol{T},\boldsymbol{W})$ with $k=\frac{c}{24}$. 

In the $\mathfrak{sl}(2)$ limit $\Phi_{W}\rightarrow 0$, 
the boundary metric reduces to $e^{\Phi_L}\, dz d\bar z$. 
By construction, the spin-$3$ field 
$\Psi$ is in FG gauge, with $\Psi_{\rho \rho \rho}=\Psi_{\rho \rho i}=0$. 
The $\rho$-expansion \eqref{phiijk} of $\Psi_{ijk}$ 
starts only at $\mathcal{O}(\rho^0)$ with 
\begin{equation}\label{FGpsiconf}
\begin{aligned}
&\psifour=
2 \, dw\, dz^2 + {\rm c.c.} \qquad\qquad
\psisix =-
dw\,dz\big(e^{-2\Phi_1+\Phi_2}d\bar{t}_1
+e^{-2\Phi_2+\Phi_1}d\bar{t}_2\big)+{\rm c.c.}\\
&\psieight =
e^{-\Phi_1-\Phi_2}\,dw\,d\bar{t}_1\,d\bar{t}_2+ {\rm c.c.}
\end{aligned}
\end{equation} 
For the expansion of $\Psi_{\rho i j}$ we find
\begin{equation}
\begin{aligned}
&\psizero_{\rho }
=-2e^{\Phi_L}\sinh(\Phi_W)dz d\bar{z} \qquad \qquad 
\psitwo_{\rho }=-\textstyle{\frac{2}{3}}\,\partial \bar{\partial}\Phi_W\,dz d\bar{z}\\
&\psifour_{\rho}=-
e^{\Phi_2-2\Phi_1}|dt_1|^2+e^{\Phi_1-2\Phi_2}|dt_2|^2
\end{aligned}
\end{equation}
We observe that there is no boundary spin-three field $\psizero_{ijk}$ in 
conformal gauge. Instead $\psi^{\scriptscriptstyle(0)}_{\rho}$
is non-zero and depends on the two Toda fields. 
Note also that the vev of the spin-3 current appears at $\mathcal{O}(\rho^0)$, 
which is a generic feature, valid for all spins and in all gauges.

Since the Toda fields are introduced via a finite PBH transformation on the 
connection \eqref{freevevslN}, further
PBH transformations are very transparent. The PBH transformation 
between two solutions labeled by $\{\boldsymbol{\Phi}_1\}$ 
and $\{\boldsymbol{\Phi}_2\}$ 
is simply
\begin{equation}
g=(g_1)^{-1}g_2
\end{equation}
The effect on a solution in conformal gauge is the shift 
of the conformal modes ($s=L,W$)
\begin{equation}\label{PBHconformal}
\Phi_s \rightarrow \Phi_s + 2\,\sigma_s 
\end{equation}
With $(\sigma_{L},\sigma_{W})\equiv (\sigma_2,\sigma_3)$ 
the boundary metric $\gz$ transforms as 
\begin{equation}\label{deltagconf}
\delta_{\sigma_2} \gz= 2\,\sigma_2 \,\gz \qquad \textrm{and} 
\qquad \delta_{\sigma_3} \gz= -\sigma_3  \, \psizero_{\rho }
\end{equation}
There is no boundary spin-3 field and the leading term of $\Psi_{\rho ij}$ 
transforms as
\begin{equation}\label{deltapsiconf}
\delta_{\sigma_2} \psizero_{\rho}= 2 \,\sigma_2 \, \psizero_{\rho } \qquad \textrm{and} 
\qquad \delta_{\sigma_3} \psizero_\rho= - 4\,\sigma_3  \, \gz
\end{equation}

Using \eqref{deltaShs} and \eqref{aconfslN} one finds that the effective action transforms 
as
\begin{equation}\label{Weylsconformal}
\begin{aligned}
\delta_{\sigma_{s}} W=\frac{t^{(s)}_0\, k}{ \pi}\,\int_{\p{\cal M}}\sigma_{s}\, 
\partial \bar{\partial }\Phi_{s}\,dz\wedge d\bar{z} 
\end{aligned}
\end{equation}
from which one reads off the $\mathcal{W}_s$ anomaly: 
\begin{equation}\label{Weylsconformal}
\begin{aligned}
\sqrt{g}\,\mathcal{A}_s=(2 \, t^{(s)}_0 \, k)\, \partial \bar{\partial }\Phi_{s}
\end{aligned}
\end{equation}
This can be integrated to
\begin{equation}
W=\sum_s {t^{(s)}_0\,k\over 4\pi}\int_{\p {\cal M}}\Phi_s\,\p\bar\p\Phi_s
\,dz\wedge d\bar z
\end{equation}
This generalizes, in a natural way, eq.\eqref{Weylconformalsl2} to the case where 
several Toda fields are present and we can view $W$ as the generalization of the 
non-local Polyakov action which has become local in conformal gauge.  
This agrees with \cite{Hull:1991sa}; but what is puzzling is that here we do not have a 
boundary spin-3 field --- $\psi^{\scriptscriptstyle(0)}_{ijk}=0$ --- but instead 
$\psi^{\scriptscriptstyle(0)}_{\rho ij}\neq 0$. 

We note that for $s=2$, using \eqref{kc}
and (\ref{gCsl3}) we find from \eqref{Weylsconformal}
\begin{equation}
\mathcal{A}_2= \frac{c}{ 6}\,\textrm{tr}\big(\gt\big)
\end{equation} 
as for pure gravity. However, except for $N=2$ 
this is not proportional to the Ricci scalar of the boundary metric.
For $N>2$, since higher-spin fields also transform under the Weyl transformation, 
${\cal A}_2$ contains contributions from the trace anomaly of the 
energy-momentum tensor and from higher-spin background fields, and it reduces to 
$T^i_{\phantom{i}i}$ only if all scalars except $\Phi_2$ are switched off 
(the $\mathfrak{sl}(2)$ limit).

\subsection{$\mu$ - gauge}

The second gauge choice is what we called the $\mu$-gauge in Chapter \ref{sec:Spin-2}.
In this gauge higher-spin sources $\{\mu_s\}$ are turned on while the conformal modes 
are set to zero. 
We explain $\mathfrak{sl}(3)$ in detail, where
in addition to $(\mu_2,\bar{\mu}_2)$, the sources for $(T,\bar T)$, 
we can also turn on $(\mu_3,\bar{\mu}_3)$, the sources for 
$(W,\bar W)$.\footnote{We use the notation
$(T,\bar T)\equiv (T_{zz},T_{\bar z\bar z})$ 
and $(W,\bar W)\equiv (W_{zzz},W_{\bar z\bar z\bar z})$.}
We will refer to the pair $(\mu_2,\mu_3)$ as 
the generalized complex structure and to $(T,W)$ as the generalized projective 
structure. 

Just as the Virasoro Ward identity (\ref{ward2}) can be obtained as the 
compatibility condition of the complex structure $\mu$ and the projective structure 
$T$, the $\mathcal{W}_3$ Ward identities can be derived from the compatibility condition 
between the generalized structures \cite{Bilal:1990wn}. 

In analogy to  (\ref{LSsl2}) and \eqref{sl2achiral} for $\mathfrak{sl}(2)$, the linear 
system for $\mathfrak{sl}(3)$ acts on a 3-vector $\Psi$ whose 
last component we denote by $\psi$. The flat 
$\mathfrak{sl}(3)$ connection that encodes the  $\mathcal{W}_3$ Ward identity is
\begin{equation}\label{sl3achiral}
\begin{aligned}
a_z&= L_{1}+\boldsymbol{q}
\qquad \textrm{and}\qquad 
a_{\bar{z}}=\boldsymbol{\mu}
+\boldsymbol{\bar{\omega}}
+ \boldsymbol{\gamma} +
\boldsymbol{\beta}
\end{aligned}
\end{equation}
The connection $a$ is in highest weight gauge: in the charge vector
\begin{equation}\label{Qsl3}
\boldsymbol{q}=-{\textstyle{1\over4}}\,{T}L_{-1}+{\textstyle{1\over4}}\,{W}W_{-2}
\end{equation}
$(T,W)$, the rescaled (by $1/k$) stress energy and spin-$3$ current, 
are along the highest weight directions
$(L_{-1}, W_{-2})$ in $a_{z}$ (the prefactors are due to the normalization 
$t^{(2)}_1=-4$ and $t^{(3)}_2=4$); in the source vector
\begin{equation}\label{Msl3}
\boldsymbol{\mu}=\mu_2 L_{1}+\mu_3 W_{2}
\end{equation}
$(\mu_2,\mu_3)$ are along the lowest weight directions $(L_1,W_2)$. 
\begin{equation}\label{ZMsl3}
\boldsymbol{\bar{\omega}}
=\bar{\omega}_2 L_0 + \bar{\omega}_3 W_0
\end{equation}
is along the zero modes and, finally, 
\begin{equation}
\boldsymbol{\gamma}=\gamma\, W_1 \qquad \textrm{and}\qquad 
\boldsymbol{\beta}=\beta_{1}\, L_{-1}+\beta_{2}\, W_{-1}+ \beta_{3}\,W_{-2}
\end{equation}
Requiring flatness of $a$ gives eight equations, six of which can be solved algebraically   
for $\boldsymbol{\bar\omega,\gamma}$, and $\boldsymbol{\beta}$: 
\bea
&\gamma= -\partial \mu_3 \qquad \qquad 
\bar{\omega}_2=- \partial \mu_2   
\qquad \qquad \bar{\omega}_3
=\textstyle{\frac{1}{2}}(\partial^2-T)\mu_3\label{flatsl3chiral}\\
\noalign{\vskip.2cm}
&\beta_1=-\textstyle{\frac{1}{2}}\partial \bar{\omega}_2} 
-\textstyle{\frac{1}{4}}{\mu_2 T}-\textstyle{\frac{1}{2}}{\mu_3 W}\,, 
\quad\beta_2=-\textstyle{\frac{1}{3}}{\partial \bar{\omega}_3}
-\textstyle{\frac{1}{4}}{\gamma T} \,, \quad 
\beta_3=\textstyle{\frac{1}{4}}{\mu_2 W}
-\textstyle{\frac{1}{8}}{\bar{\omega}_3 T-\textstyle{\frac{1}{4}}{\partial \beta_2}
\nonumber
\eea
The remaining two equations are the two $\mathcal{W}_3$ Ward identities:\footnote{They
already appeared e.g. in \cite{Bilal:1990wn} and recently in a context similar to the 
one of this paper, in \cite{deBoer-Jottar, Indians}.}
\begin{equation}\label{WIsl32}
\begin{aligned}
\big(\bar{\partial}-\mu_2\,\partial -2\,\partial\mu_2\big)T &=-2\,\partial^3\mu_2 
+\big(2\, \mu_3\, \partial +3\, \partial \mu_3\big)W
\end{aligned}
\end{equation}
for $T$ and 
\begin{equation}\label{WIsl33}
\begin{aligned}
&\big(\bar{\partial}-\mu_2\, \partial -3\,\partial \mu_2\big)W\\
\noalign{\vskip.2cm}
&\qquad=\textstyle{\frac{1}{6}}\,\partial^5 \mu_3
-\big(\textstyle{\frac{1}{6}}\mu_3 \,\partial^3 
+\textstyle{\frac{3}{4}}\,\partial \mu_3\, \partial^2 
+\textstyle{\frac{5}{4}}\,\partial^2 \mu_3\, \partial 
+\textstyle{\frac{5}{6}}\,\partial^3 \mu_3)\,T
+\textstyle{\frac{1}{3}}(\mu_3\, \partial +2\, \partial \mu_3\big)\,T^2
\end{aligned}
\end{equation}
for $W$. Together they encode the $\mathcal{W}_3$-algebra. 
The first term on the r.h.s. of each identity is the anomaly.  The $\mathfrak{sl}(3)$ analogues of the pair of equations (\ref{eqpsi}) on $\psi$ 
are a third-order holomorphic equation
\begin{equation}\label{eqpsi3z}
\left(\partial^3- T \partial -\textstyle{\frac{1}{2}}\partial T +W\right)\,\psi=0
\end{equation}
and
\begin{equation}\label{eqpsi3zb}
\Big[\bar\p+\mu_3\,\p^2-(\mu_2+{1\over 2}\p\mu_3)\p+\p\mu_2+
\textstyle{1\over6}\p^2\mu_3-{2\over3}\mu_3\,T
\Big]\psi=0
\end{equation}
The Ward identities \eqref{WIsl32} and \eqref{WIsl33} can be 
obtained as compatibility conditions between (\ref{eqpsi3z}) and (\ref{eqpsi3zb}).

As in the case of $\mathfrak{sl}(2)$, these Ward identities can also be 
understood directly as compatibility conditions of the generalized projective and 
complex structures. To explain this, we write the connection, which is pure gauge, 
in the form 
\begin{equation}
a=g^{-1} d\,g
\end{equation}
We make a Gauss decomposition Ansatz for $g$ where, for convenience
we use the Chevalley basis
\begin{equation}\label{basis}
g=e^{w_{2}W_2+e_{+}E_{+}+f_{+}F_{+}}\, e^{-\phi_{1}H^{(1)}-\phi_{2}H^{(2)}} \, 
e^{w_{-2}W_{-2}+e_{-}E_{-}+f_{-}F_{-}}
\end{equation}
Requiring $a$ to be in highest weight gauge gives  
\begin{equation}\label{dbw2}
\partial w_2={\textstyle\frac{1}{4}}\big(f_{+}\,\partial e_{+}-e_{+}\,\partial f_{+}\big)\,,
\end{equation}
relates the $+1$ modes to the zero mode:
\begin{equation}\label{fw0}
\partial e_{+} =-\sqrt{2}\,  e^{2 \phi_1 -\phi_2} \qquad \textrm{and}\qquad 
\partial f_{+} =-\sqrt{2}\, e^{2 \phi_2 -\phi_1}
\end{equation}
and solves all negatives modes in terms of the zero modes
\begin{equation}
e_{-}=\frac{1}{\sqrt{2}}\partial \phi_{1}\,,\qquad f_{-}
=\frac{1}{\sqrt{2}}\partial \phi_{2}\,, \qquad w_{-2}
=\frac{1}{8}\big[(\partial \phi_1)^2-\partial^2 \phi_1\big]
-\frac{1}{8}\big[(\partial \phi_2)^2-\partial^2 \phi_2\big]\,.
\end{equation} 
$(T,W)$ can be read off from the $(L_{-1},W_{-2})$ direction of $a_z$; 
they are the negative 
of the two conserved charges \eqref{chargeToda3} of the Toda theory 
with Toda fields $(\phi_1,\phi_2)$. Using (\ref{fw0}) 
we can rewrite $(T,W)$ in terms of the (holomorphic derivatives of 
the) two positive modes in the simple root directions, 
$(e_{+},f_{+})$. These are the generalizations of the 
Schwarzian derivative \eqref{Schwarzian} to $\mathfrak{sl}(3)$:\footnote{
We note that this result has been found long time ago, see e.g. \cite{Marshakov:1989ca}.
Finite $W_3$-transformations of
$T$ and $W$, which reduce to the Schwarzian derivative 
when the initial charges are set to zero, as was the case considered here, 
were found in \cite{Gomis:1994rz}. }
\begin{eqnarray}
\label{sl3Schwarz}
T&=&-\left\{ {\p^3 e\over\p e}
-{4\over 3}\left({\p^2 e\over\p e}\right)^2\right\}
+{1\over 6}{\p^2 e\over\p e}{\p^2 f\over\p f}+\big(e\leftrightarrow f\big)\\
\noalign{\vskip.2cm}
W&=&-{1\over 6}\left\{{\p^4 e\over\p e}-5{\p^3 e\over \p e}{\p^2 e\over \p e}
+{40\over 9}\left({\p^2 e\over \p e}\right)^3\right\}
+{1\over 6}{\p^3 e\over\p e}{\p^2 f\over\p f}+{5\over18}{\p^2 e\over\p e}
\left({\p^2 f\over\p f}\right)^2 -(e\leftrightarrow f)
\nonumber\end{eqnarray}
Here and later  we drop the subscripts in $(e_+,f_+)$ 
for better readability.
When expressed in terms of the zero modes $\phi_i$, these are simply the 
Toda charges. 
For $f=e=u$, $W=0$ and $T$ reduces to $T=-2\{u,z\}$, i.e. the stress energy 
$\boldsymbol{T}$ reduces to the $\mathfrak{sl}(2)$ result 
$\boldsymbol{T}=-\frac{c}{12}\{u,z\}$, cf. \eqref{Schwarzian}.
For infinitesimal transformations parametrized by $\epsilon(z)$ and $\p\eta(z)$, 
\begin{equation}\label{infinitesimal}
e=z+\epsilon(z)-\textstyle{\frac{1}{2}}\p\eta(z)\,,\qquad
f=z+\epsilon(z)+\textstyle{\frac{1}{2}}\p\eta(z)
\end{equation}
one finds
\begin{equation}
T=-2\,\p^3\epsilon\,,\qquad W={1\over6}\,\p^5\eta
\end{equation}
Note that $\p\eta$ rather than $\eta$ appears in \eqref{infinitesimal} so that for 
infinitesimal transformations all functions in \eqref{basis} 
can be expressed in terns of $\xi$, $\eta$ and their (holomorphic) derivatives.

The generalization of the Beltrami equation (\ref{isothe}) is obtained by solving 
$(\mu_2,\mu_3)$ from $a$: $\mu_2$ and $\mu_3$ are algebraic functions of the 
positive modes $(w_{2}, e_+, f_+)$ and their derivatives (both holomorphic and 
anti-holomorphic). In these functions  $w_{2}$ only appears in the form of 
$\bar{\partial}w_{2}$, then applying $\partial$ on it and using (\ref{dbw2}) 
we can replace it in favor of $(e_+,f_+)$, and obtain two equations on $(e_+,f_+)$:
\begin{equation}\label{sl3Beltrami}
\begin{aligned}
\left[\bar{\partial}-\Big(\mu_2-{\frac{1}{2}}\partial \mu_3 
-{\frac{2}{3}}\mu_3\,\frac{\partial^2 f}{\partial f}\,\Big)\,\partial 
+{\frac{1}{3}}\mu_3\,\partial^2 \, \right] e =0 \\
\left[\bar{\partial}-\Big(\mu_2+{\frac{1}{2}}\partial \mu_3 
+{\frac{2}{3}}\mu_3\,\frac{\partial^2 e}{\partial e}\,\Big)\,\partial 
-{\frac{1}{3}}\mu_3\,\partial^2 \,\right] f =0
\end{aligned}
\end{equation}
These are the $\mathfrak{sl}(3)$ analogues of the Beltrami equation 
\eqref{isothe}. They agree with eqs. (4.10) in \cite{Bilal:1990wn} 
after the redefinition $\mu_2\to\mu_2-{1\over2}\p\mu_3$.
The Ward identities (\ref{WIsl32}) and (\ref{WIsl33}) can be obtained as 
the compatibility conditions between them and the generalized Schwarzian 
derivatives \eqref{sl3Schwarz}. As a consistency check one can verify 
\eqref{sl3Beltrami} by using \eqref{epfp} below, \eqref{eqpsi3z} and \eqref{eqpsi3zb}.

Eqs. \eqref{sl3Beltrami} can be simplified to yield one algebraic equation relating 
$\mu_2$ and $\mu_3$ and one first-order differential equation for $\mu_3$:
\begin{equation}\label{Beltrami32}
\begin{aligned}
\mu_2&=-\frac{1}{6}\,\mu_3 \,\left(\frac{\partial^2 e}{\partial e}
-\frac{\partial^2 f}{\partial f}\right)+\frac{1}{2} \,\left(\frac{\bar{\partial} e}
{\partial e}+\frac{\bar{\partial} f}{\partial f}\right)\\
\noalign{\vskip.2cm}
\partial\mu_3&=-\phantom{\frac{1}{6}\,}\mu_3 \,\left(\frac{\partial^2 e}{\partial e}
+\frac{\partial^2 f}{\partial f}\right)- \left(\frac{\bar{\partial} e}{\partial e}
-\frac{\bar{\partial} f}{\partial f}\right)
\end{aligned}
\end{equation}

The Schwarzian derivative \eqref{Schwarzian} is invariant under M\"obius 
transformation $u \rightarrow \frac{a \, u+b}{c \, u +d}$, which is a 
non-linear realization of $\textrm{SL}(2)$. To obtain the non-linear realization 
of $\textrm{SL}(3)$ action on $(e_+, f_+)$ which leaves \eqref{sl3Schwarz} invariant, we 
use the fact that $e_+$ and $f_+$ can be related to the two ratios 
\begin{equation}\label{u12}
u_1\equiv \frac{\psi_1}{\psi_3} \qquad \textrm{and} 
\qquad  u_2\equiv \frac{\psi_2}{\psi_3}
\end{equation} 
of any three independent solutions $\psi_i$ of (\ref{eqpsi3z}) and (\ref{eqpsi3zb}) 
as \cite{Bilal:1990wn} 
\begin{equation}\label{epfp}
f=u_2\qquad 
\textrm{and} \qquad  e=\frac{\p u_1}{\p u_2}
\end{equation}
Since any linear combination of the three solution vectors $\Psi_i$ of the linear 
system (\ref{LSsl2}) with $a$ given in (\ref{sl3achiral}) is still a solution,  
the generalized Schwarzian derivatives (\ref{sl3Schwarz}) are invariant under 
the following  $\textrm{GL}(3)$ action 
\begin{equation}\label{gl3}
u_1 \rightarrow \frac{c_{11}\,u_1+c_{12}\,u_2+c_{13}}{c_{31}\,u_1+c_{32}\,u_2+c_{33}} 
\qquad \textrm{and} \qquad 
u_2 \rightarrow \frac{c_{21}\,u_1+c_{22}\,u_2+c_{23}}{c_{31}\,u_1+c_{32}\,u_2+c_{33}} 
\end{equation}
One also checks that the r.h.s.'s of \eqref{sl3Schwarz} vanish for 
\begin{equation}\label{uz}
u_1={c_{11}\,{z^2\over 2}+c_{12}\,z+c_{13}
\over c_{31}\,{z^2\over2}+c_{32}\,z+c_{33}}\,,\qquad
u_2={c_{21}\,{z^2\over2}+c_{22}\,z+c_{23}
\over c_{31}\,{z^2\over2}+c_{32}\,z+c_{33}}
\end{equation}
where $\{c_{ij}\}\in {\rm GL}(3)$. For infinitesimal transformations
with $c_{ij}=\delta_{ij}+\gamma_{ij}$, 
$u_1$ and $u_2$ lead to quadratic and quartic polynomials (in $z$) for 
$\epsilon$ and $\eta$ (cf. \eqref{infinitesimal}); for this the factors $\frac{1}{2}$ in (\ref{uz}) are needed.
Recall that infinitesimal M\"obius 
transformations are quadratic polynomials, which reflects the fact that 
normalizable holomorphic vector fields $\xi^i$ which
generate infinitesimal diffeomorphisms on the Poincar\'e sphere grow at most as 
$z^2$ for large $z$. `Generalized diffeomorphisms' are 
generated by symmetric tensors $\xi^{ij}$ which are normalizable as long as they 
do not grow faster than $z^4$. Put differently, there are three (five) $c$-ghost
zero modes on the sphere for a $\lambda=2$ ($\lambda=3$) $(b,c)$ ghost system
(and none for the $b$-ghost).

The generalizations of the construction that gives \eqref{sl3Schwarz} and 
\eqref{sl3Beltrami} to 
$SL(N)$ with $N>3$, and presumably also to other other
classical groups, are straightforward. Some results for $SL(N)$ are presented 
in Appendix \ref{sec:SL(N)}.

After this excursion to generalized complex and projective structures, 
we now combine the flat connection \eqref{sl3achiral} with $\tilde a=-a^\dagger$. 
Clearly $(a,\tilde a)$ is not in FG gauge: their zero-mode parts do not match.  
The task is to modify them in such a way that they are in FG gauge  while 
still encoding, via the flatness conditions, the Ward identities. In doing so we
want to preserve the following properties:  
the only positive mode in $a_z$ is along $L_{+1}$ as this captures the AdS$_3$ 
vacuum and the charges are along the highest weight directions. 
This can be accomplished by a gauge transformation of \eqref{sl3achiral} 
such that 
\begin{equation}
L_{1} \rightarrow L_{1}-\boldsymbol{\omega} + \textrm{negative modes} 
\end{equation}
with
\begin{equation}\label{omega}
\boldsymbol{\omega}=\boldsymbol{\bar{\omega}}^{\dagger}=\omega_2 L_0 +\omega_3 W_0
\end{equation}
A moment's thought reveals that this is solved by 
\begin{equation}\label{CtoNCsl3}
g=M^{-1}
\end{equation}
where $M=M(\boldsymbol{\omega})$ in the Miura transformation given explicitly in 
\eqref{Miurasl3}, which satisfies (cf. \eqref{Miura})
\begin{equation}
M(L_1-\mathcal{Q}(\boldsymbol{\omega}))M^{-1}-\partial M M^{-1}=L_1-\boldsymbol{\omega}
\end{equation}
with $\mathcal{Q}$ as in \eqref{chargeToda3}. The gauge transformed connection 
has the structure
\begin{equation}\label{aNC}
\begin{aligned}
&a_z = L_1 -\boldsymbol{\omega} + \boldsymbol{q} \qquad \qquad \qquad
a_{\bar{z}} = \boldsymbol{\mu}+\boldsymbol{\bar{\omega}}+\boldsymbol{\beta} 
+\boldsymbol{\gamma}\\
\noalign{\vskip.2cm}
&\tilde{a}_z =\boldsymbol{\bar{\mu}}+\boldsymbol{\omega}+\boldsymbol{\bar{\beta}} 
+\boldsymbol{\bar{\gamma}}\qquad \qquad~
\tilde{a}_{\bar{z}} = L_{-1} +\boldsymbol{\bar{\omega}} + \boldsymbol{\bar{q}}
\end{aligned}
\end{equation}
where 
\begin{equation}
\boldsymbol{\bar{q}}=-\boldsymbol{q}^{\dagger}\,,\qquad 
\boldsymbol{\bar{\mu}}=-\boldsymbol{\mu}^{\dagger}\,, \qquad
\boldsymbol{\bar{\gamma}}=-\boldsymbol{\gamma}^{\dagger}\,,\qquad
\boldsymbol{\bar{\beta}}=-\boldsymbol{\beta}^{\dagger}
\end{equation} 
We emphasize that given the same $(\mu_2, \mu_3)$, the remaining variables 
in (\ref{sl3achiral}) and (\ref{aNC}) take different values.

The effect of the gauge transformation (\ref{CtoNCsl3}) on the charge vector is 
\begin{equation}
\boldsymbol{q} \rightarrow M(\boldsymbol{q}+\mathcal{Q}(\boldsymbol{\omega}))M^{-1}\\
\end{equation}
This gives 
\begin{equation}\label{CtoNCQ}
\begin{aligned}
T&\rightarrow T+ \mathcal{Q}_2(\boldsymbol{\omega})  \qquad \textrm{and}\qquad 
W\rightarrow [W+ \mathcal{Q}_3(\boldsymbol{\omega})]
-{\textstyle\frac{1}{3}}{\omega_3} [T +\mathcal{Q}_2(\boldsymbol{\omega}) ] \\
\end{aligned}
\end{equation}
Similarly, the sources are shifted 
\begin{equation}\label{CtoNCmu}
\mu_2 \rightarrow \mu_2+\frac{\omega_3}{3}\mu_3
\qquad \textrm{and}\qquad \mu_3 \rightarrow \mu_3
\end{equation}
The charge vector therefore transforms as
\begin{equation}
\mu_2 T + \mu_3 W \rightarrow \mu_2 (T+\mathcal{Q}_2) + \mu_3 (W+\mathcal{Q}_3)
\end{equation}
Applying the gauge transformation (\ref{CtoNCsl3}) we see that the six algebraic 
equations \eqref{flatsl3chiral} in the flatness condition of $a$ change to
\begin{equation}\label{sl3aflatness}
\begin{aligned}
&\gamma= -(\partial +2 \omega_2)\mu_3\,,\\
\noalign{\vskip.2cm}
&\mu_2\,\omega_2+\bar{\omega}_2=-\partial \mu_2
-{\textstyle\frac{1}{2}}\gamma \,\omega_3\,,\qquad 
\mu_2 \,\omega_3+\bar{\omega}_3=
-{\textstyle\frac{1}{2}}\big(\mu_3 T+(\p+\omega_2)\gamma\big)\\
\noalign{\vskip.2cm}
&\beta_1=-{\textstyle\frac{1}{2}}\big({\p\bar{\omega}_2+\bar{\p}\omega_2}\big) 
-{\textstyle\frac{1}{4}}{\mu_2 T}
-{\textstyle\frac{1}{2}}\mu_3 W\,, \qquad
\beta_2=-{\textstyle\frac{1}{3}}\big({\partial \bar{\omega}_3
+\bar{\partial}\omega_3}\big)
-{\textstyle\frac{1}{4}}{\gamma T}\,, \qquad \\
\noalign{\vskip.2cm}
&\beta_3={\textstyle\frac{1}{4}}{\mu_2 W}
-{\textstyle\frac{1}{8}}{\bar{\omega}_3 T}
-{\textstyle\frac{1}{4}}{\partial \beta_2} 
+\left({\textstyle\frac{1}{4}}{\omega_2 \beta_2}
+{\textstyle\frac{1}{2}}{\omega_3 \beta_1 }\right)
\end{aligned}
\end{equation}
and the complex conjugate set of conditions from flatness of $\tilde{ a}$.  
Note that we can no longer express the zero modes algebraically in terms 
of the sources and their derivatives. This is in contrast to the 
situation for $\mathfrak{sl}(2)$, which we discussed in Section 
\ref{sec:sl2-mugauge}.

The Ward identities take the same form as the chiral Ward identities (\ref{WIsl32}) 
and (\ref{WIsl33}) with the replacements
\begin{equation}\label{NCtoCQmu}
\begin{aligned}
&\mu_2 \rightarrow \mu_2 -{\textstyle\frac{1}{3}}\,{\omega_3}\,\mu_3\\
\noalign{\vskip.2cm}
&T\rightarrow T - \mathcal{Q}_2(\boldsymbol{\omega})\qquad \textrm{and}\qquad 
W\rightarrow W-\mathcal{Q}_3(\boldsymbol{\omega})
+{\textstyle\frac{1}{3}}\,{\omega_3}\,T 
\end{aligned}
\end{equation}
which are the inverse transformation of (\ref{CtoNCQ}) and (\ref{CtoNCmu}); 
$\bar T$ and $\bar W$ satisfy the same identities with $\mu\to\bar\mu$, etc. 

The dreibein, the metric, and the spin-three field and their FG expansions are now
easily computed and they follow the general pattern outlined before. 
In particular, the metric is strongly backreacted as long as $\mu_3\bar\mu_3\neq 0$.  
To avoid this we can set $\bar{\mu}_3=0$, and obtain a configuration with boundary 
metric at $\mathcal{O}(\rho^{-2})$ and spin-$3$ field at $\mathcal{O}(\rho^{-4})$:
\begin{equation}
\begin{aligned}
\gz=|dz+\mu_2\, d\bar{z}|^2 \qquad
\hbox{and}\qquad\psizero=\mu_3 \, d\bar{z} (d\bar{z}+\bar{\mu}_2\, dz)^2
\end{aligned}
\end{equation} 
The flatness condition of this configuration follows by setting $\bar{\mu}_3=0$ 
in (\ref{sl3aflatness}) and its complex conjugate. We can see that even with 
$\bar{\mu}_3=0$ the zero modes still cannot be solved algebraically in terms 
of the sources and their derivatives. 

This is not a problem in itself. However, to simplify the computation (without losing 
real content)  we consider 
\begin{equation}
\bar{\mu}_2= \bar{\mu}_3= 0  
\end{equation}
which generalizes the lightcone gauge  for $\mathfrak{sl}(2)$ (\ref{sl2aNCchiral}) to $\mathfrak{sl}(3)$. The connections \eqref{aNC} simplify to
\begin{equation}\label{aNCchiral}
\begin{aligned}
&a_z = L_1 + \boldsymbol{q} \qquad \qquad 
a_{\bar{z}} = \boldsymbol{\mu}+\boldsymbol{\bar{\omega}}+\boldsymbol{\beta} 
+\boldsymbol{\gamma}\\
&\tilde{a}_z =\boldsymbol{\bar{\beta}} 
\,\,\qquad \qquad  \qquad 
\tilde{a}_{\bar{z}} = L_{-1} +\boldsymbol{\bar{\omega}} + \boldsymbol{\bar{q}}
\end{aligned}
\end{equation}
In particular $a$  has reduced to (\ref{sl3achiral}), from which we 
derived the $\mathcal{W}_3$ Ward identities.  Therefore the parameters in 
$(\boldsymbol{\bar{\omega}},\boldsymbol{\beta},\boldsymbol{\gamma})$ are as in 
(\ref{flatsl3chiral}).   
However $\tilde{a}$ is no longer $- a^{\dagger}$. In particular 
$\boldsymbol{\bar{\beta}}
=\bar{\beta}_1L_1+\bar{\beta}_2W_1 -\bar{\beta}_3W_2\neq\boldsymbol{\beta}^\dagger$,
but instead
\begin{equation}
\bar{\beta}_1=-\textstyle{\frac{1}{2}}\partial \bar{\omega}_2 \qquad \bar{\beta}_2
=-\textstyle{\frac{1}{3}}\partial \bar{\omega}_3 \qquad \bar{\beta}_3
=\textstyle{\frac{1}{12}}\partial \bar{\partial} \bar{\omega}_3 
-\textstyle{\frac{1}{12}}\bar{\omega}_2\partial\bar{\omega}_3
-\textstyle{\frac{1}{4}}\bar{\omega}_3\partial\bar{\omega}_2
\end{equation}
With (\ref{aNCchiral}) we can compute the $\rho$-expansion of the dreibein, which 
starts with $\eminusone$ at $\mathcal{O}(\rho^{-2})$  due to the presence of the 
spin-$3$ source. We give the
leading terms: 
\begin{equation}\label{drei0}
\begin{aligned}
&\eminusone=\textstyle{\frac{1}{2}}\,
\mu_3 \, W_{2}\,d\bar{z} \,,\qquad \ez=-L_0 \, d\rho
+\frac{1}{2}\left[L_{1}\, (dz+\mu_2 \,d\bar{z})
- \partial \mu_3\, W_{1}\,d\bar{z}-L_{-1}\,d\bar{z}\right] 
\end{aligned}
\end{equation}

The $\mathcal{O}(\rho^{-4})$ term of the bulk metric is now absent but the leading 
term of the bulk spin-$3$ field remains. 
The boundary metric and spin-$3$ field are
\begin{equation}
\begin{aligned}
\gz=(dz+\mu_2\, d\bar{z}) d\bar{z} \qquad
\hbox{and}\qquad\psizero=\mu_3 \, d\bar{z}^3
\end{aligned}
\end{equation} 
With only a chiral spin-$3$ source, the bulk energy-momentum tensor is such 
that it does not modify the asymptotic behavior of the bulk metric. 
The boundary value of $\Psi_{\rho i j}$ is also present (as in the conformal gauge):
\begin{equation}
\psizero_{\rho}=(\partial \mu_3) d\bar{z}^2
\end{equation}
and vanishes only when $\partial \mu_3=0$. 

Under PBH transformations the boundary metric $\gz$ transforms 
as\footnote{The PBH transformation preserves the FG gauge but does not 
preserve \eqref{aNCchiral}.} 
\begin{equation}\label{deltagchiralmu}
\begin{aligned}
\delta_{\sigma_2}\gz=2\,\sigma_2 \,  \gz  \qquad\textrm{and}\qquad
\delta_{\sigma_3}\gz 
= -\sigma_3 \,\psizero_{\rho} +\frac{2}{3}\,\partial \sigma_3\,\mu_3
\,d\bar{z}^2
\end{aligned}
\end{equation}
while for the boundary spin-$3$ field $\psizero$ one finds
\begin{equation}\label{PBH3psi}
\delta_{\sigma_2}\psizero=4\,\sigma_2\,\psizero\qquad \textrm{and}\qquad 
\delta_{\sigma_3} \psizero =0
\end{equation}
In contrast to (\ref{WWeylgeneral}), where the spin-3 source has a strong 
back-reaction, the boundary spin-$3$ field now is invariant under the  
spin-$3$ Weyl transformation, whereas the boundary metric transforms.
Finally, the $\psi^{\scriptscriptstyle{(0)}}_{\rho i j}$ component  transforms as 
\begin{equation}\label{deltapsirhochiralmu}
\delta_{\sigma_2}\psizero_{\rho}=2\,\sigma_2\,\psizero_{\rho}+4\,\partial 
\sigma_2\,\mu_3\,d\bar{z}^2 
\qquad\textrm{and}\qquad\delta_{\sigma_3}\psizero_{\rho}=-4\,\sigma_3\gz
\end{equation}

From (\ref{deltaShs}) we find for the 
${\cal W}$-Weyl variations of the effective action
\begin{equation}\label{PBHsl2NC2}
\delta_{\sigma_s} W
=\frac{t_0^{(s)}k}{\pi}\int_{\partial\mathcal{M}}
\sigma_s\,(\partial\bar\omega_s)\,dz\wedge d\bar{z}
\end{equation}
The interpretation of this is not obvious. For $s=2$, using \eqref{flatsl3chiral}, 
we see that this agrees with \eqref{PBHsl2NCchiral} and there is no contribution from the 
spin-3 field. For $s=3$, using once again \eqref{flatsl3chiral} we find
\begin{equation}\label{PBHsl3NC3}
\delta_{\sigma_3} W
=\frac{k}{3\pi} \int_{\partial\mathcal{M}}d^2z 
\,\sigma_3\,\partial\big[(\partial^2-T)\mu_3\big]
\end{equation}
The appearance of $T$ is at first surprising. On second thought it might be expected
because the boundary metric, the source for $T$, transforms under $\sigma_3$. 
The details of how \eqref{PBHsl3NC3} results are, however, not clear to us.  

\section{Conclusions}\label{sec:}

Higher-spin theories in three dimensions coupled to gravity with a 
negative cosmological constant are most easily described in terms of Chern-Simons 
theories. They are dual to conformal field theories 
with ${\cal W}$-symmetry. These two-dimensional field theories 
can be coupled to external sources; the source for the conserved spin-$s$ current is a spin-$s$ gauge field.     
The usual way to describe them holographically 
is in terms of metric-like bulk fields
whose boundary configurations can be identified as the sources. This is well known and understood for pure gravity, where a
Chern-Simons description as well as a metric one are known. 
For the higher-spin generalizations the situation is considerably more 
difficult, as their formulation in the bulk is not known in terms of the 
metric-like fields. They are derived from the connections and consequently 
also the boundary data, i.e. the sources and vev's of the dual 
field theory on the boundary, have to be encoded in the connection. 

We have addressed these issues by first translating  
results known for the case of pure gravity into the Chern-Simons language 
in such a way that a generalization to the higher-rank, i.e. higher-spin 
case, is immediate. This concerns the Fefferman-Graham gauge, the 
action of the residual PBH transformations, and different gauge choices
for the boundary theory, such as conformal gauge and light-cone gauge.   
While the generalizations are straightforward, the interpretation of the results 
they lead to are not. One problem is to find configurations 
that, on the one hand, have no strong back-reaction on the metric (i.e.  
the radius of the asymptotic AdS does not change) and on the 
other, incorporate sources for the metric and the higher-spin fields.
For instance, in conformal gauge, which is parametrized by $\mathfrak{sl}(N)$ 
Toda fields, the spin-$s$ source for $s>2$ does not appear in the expected way. 
Another problem which we encountered was the interpretation of the 
variation of the effective action, which can be computed using 
holography. Only in the conformal gauge did we succeed to connect 
to results which were previously derived in the ${\cal W}$-gravity literature. 
The situation in the light-cone gauge was much less clear. However, the analysis 
of this gauge leads us to a natural generalization of the Schwarzian derivative 
from $\mathfrak{sl}(2)$ to $\mathfrak{sl}(3)$ and beyond.  

It would be interesting to explore the open issues discussed here further. 
The Fefferman-Graham gauge we have used throughout the paper relies on a particular choice of the radial 
component of the connections (\ref{ArhoAtrho}) and the matching of the zero-modes between the two connections (\ref{FGslN}). Although 
these choices are natural ---  
one reason being that they have a straightforward $\mathfrak{sl}(2)$ (i.e. pure gravity) 
limit --- their appropriateness in the higher-spin theory might be questioned. 
We hope to address some of the open problems in the 
near future. 

\section*{Acknowledgments}

We thank M. Banados, J. de Boer, S. Campoleoni, S. Fredenhagen, M. Gaberdiel, 
J. Jottar, Y. Korovin, R. Manvelyan, 
A. Perez, and Z. Skvortsov 
for helpful discussions. WL is supported by the European Research Council under the European Union's Seventh Framework Programme (FP7/2007-2013), ERC Consolidator Grant Agreement ERC-2013-CoG-615443: SPiN (Symmetry Principles in Nature).

\begin{appendix}

\section{Conventions}\label{sec:conventions}
\subsection{$\mathfrak{sl}(N)$ }

In this appendix we give details of the two bases for 
$\mathfrak{sl}(N)$ which we use and also fix some notation.  

In the Chevalley basis there is a triplet of generators 
$\lbrace H^{(i)},E^{(i)}_+,E^{(i)}_-\rbrace$ 
($E^{(i)}_-=(E^{(i)}_+)^\dagger$) for each simple root 
$\a_i,\,i=1,\dots,N-1$. The commutation relations are 
\begin{equation}\label{HEij}
[H^{(i)},E^{(j)}_{\pm}]=\pm K_{ji} E^{(j)}_{\pm} \qquad \qquad  
[E^{(i)}_{+},E^{(j)}_{-}]= \delta^{ij}H^{(i)}
\end{equation}
$K_{ij}$ is the Cartan matrix of $\mathfrak{sl}(N)$ 
\begin{equation}
K_{ij}=2\delta_{ij}-\delta_{i,j+1}-\delta_{i+1,j} =\textrm{tr}[H^{(i)}H^{(j)}]=2 
\frac{\langle\alpha_i , \alpha_j\rangle}{\langle\alpha_j , \alpha_j\rangle}
\end{equation}
$H^{(i)}$ are the Cartan generators 
\begin{equation}
H^{(i)}={\cal E}_{ii}-{\cal E}_{i+1,i+1} \qquad \qquad i=1,\dots, N-1
\end{equation}
and $E^{(i)}_\pm$ are raising and lowering operators
\be
E^{(i)}_+={\cal E}_{i,i+1}\qquad\qquad E^{(i)}_-={\cal E}_{i+1,i}
\ee
${\cal E}_{ij}$ is the matrix unit with 
matrix elements $({\cal E}_{ij})_{kl}=\delta_{ik}\delta_{jl}$, 
i.e. the only non-zero matrix element is a one in the $(ij)$ position. 
The remaining generators are obtained by the commutators 
$[E^{(i)}_\pm,E^{(j)}_\pm]$. Altogether they are 
${\cal E}_{ij},i\neq j$ and the $H^{(i)}$.

The Chevalley basis is often used when discussing Toda systems. There 
is a second basis which is adapted to the study of $\mathfrak{sl}(N)$ Chern-Simons 
theory as a gravitational theory. Since the gravity sector corresponds an 
$\mathfrak{sl}(2)$ algebra, we first need to choose a subalgebra 
embedding $\mathfrak{sl}(2)\hookrightarrow\mathfrak{sl}(N)$
\begin{equation}
[L_{m},L_{n}]=(m-n)L_{m+n} \qquad \qquad m,n=-1,0,1
\end{equation}
as the gravitational subsector. We then decompose the adjoint representation
of $\mathfrak{sl}(N)$ 
in irreducible representations $\{W^{(s)}_{m} \}$ of this gravity 
$\mathfrak{sl}(2)$, where the spin $s$ is integer or half-integer and 
$m\in\lbrace -s+1,\dots,s-1\rbrace$ is the magnetic quantum number:
\begin{eqnarray}\label{Vprimary}
[L_m,W^{(s)}_n] &=& [(s-1) m-n] W^{(s)}_{m+n}
\end{eqnarray}
(We sometimes write $W^{(2)}_m=L_m$.)

There are inequivalent embeddings $\mathfrak{sl}(2)\hookrightarrow\mathfrak{sl}(N)$ 
and they differ by the spectrum of spins. 
There is always one, the principal embedding, with the property that 
there is one spin-$s$ field for each $s=2,\dots,N$. 
In this paper we focus on the principal embedding.

In terms of the Chevalley basis, the $\mathfrak{sl}(2)$ for the principal embedding is 
\begin{equation}\label{principalsl2}
L_0 \equiv \frac{1}{2}\sum^{N-1}_{i=1}k_i\, H^{(i)}\qquad\qquad L_{\pm 1} 
\equiv\mp\sum^{N-1}_{i=1}\sqrt{k_i} E^{(i)}_{\mp}
\end{equation}
where $k_i= 2\sum_{j}(K^{-1})_{ij}$ are the heights of the fundamental weights. 
Note the switch of the $\pm$ in passing from $E_{\mp }$ to $L_{\pm 1}$.
Then starting from the  $\mathfrak{sl}(2)$ generators $\{L_{0}, L_{\pm}\}$, 
the higher-spin generators $W^{(s)}_{m}$ are 
\begin{equation}\label{Wexplicit}
W^{(s)}_m=(-1)^{s-m-1}\frac{(s+m-1)!}{(2s-2)!}(\textrm{adj}_{L_{-1}})^{s-m-1}(L_1)^{s-1}  
\end{equation}
with the adjoint action defined as $\textrm{adj}_A B =[A,B]$.
(\ref{Wexplicit}), together with $\left(L_{m}\right)^{\dagger}=(-1)^m L_{-m}
$, implies 
\begin{equation}
(W^{(s)}_{m})^{\dagger}=(-1)^m W^{(s)}_{-m} \qquad  s=2,\ldots,N
\end{equation}
The thus constructed basis is orthogonal but not normalized:
\begin{equation}\label{orthor}
\textrm{tr}[W^{(s)}_mW^{(t)}_n]=\delta^{st} \delta^{\phantom{(s)}}_{m,-n}t^{(s)}_m 
\qquad \textrm{with}\qquad t^{(s)}_{m}\equiv \textrm{tr}[W^{(s)}_m W^{(s)}_{-m}]
\end{equation}

\subsection{$\mathfrak{sl}(2)$ }

The height is $k_1=2$, from which we get the $\mathfrak{sl}(2)$ algebra 
\eqref{principalsl2}
\begin{equation}
L_0=\begin{pmatrix}
\frac{1}{2} & 0\\
0& -\frac{1}{2}
\end{pmatrix}\qquad L_1=\begin{pmatrix}
0 & 0\\
-1& 0
\end{pmatrix}\qquad L_{-1}=\begin{pmatrix}
0 & 1\\
0& 0\end{pmatrix}
\end{equation}
The normalization factors (\ref{orthor}) are 
\begin{equation}
t^{(2)}_0=\textrm{tr}[(L_0)^2]
=\frac{1}{2} \qquad \qquad t^{(2)}_1
= \textrm{tr}[L_{1}L_{-1}]
=-1
\end{equation}
Define the combinations 
\begin{equation}
J_0={1\over 2}\big(L_1+L_{-1}\big)\,,\qquad
J_1={1\over2}\big(L_1-L_{-1}\big)\,,\qquad
J_2=L_0
\end{equation}
They satisfy 
\begin{equation}
[J_a,J_b]=\epsilon_{ab}{}^c J_c\,,\qquad
{\rm Tr}(J_a J_b)={1\over2}\eta_{ab} 
\end{equation}
where $\eta_{ab}={\rm diag}(-1,+1,+1)$ is used to raise and lower indices. 

\subsection{$\mathfrak{sl}(3)$ }

The height vector is $\vec{k}=(2,2)$, from which we get via \eqref{principalsl2}
\begin{equation}\label{sl3repL}
L_0=\begin{pmatrix}
1 & 0 &0\\
0 & 0 &0\\
 0& 0 &-1
\end{pmatrix}\qquad L_1=\begin{pmatrix}
0 & 0 &0\\
 -\sqrt{2}& 0 &0\\
 0& -\sqrt{2} &0
\end{pmatrix}\qquad L_{-1}=\begin{pmatrix}
0 & \sqrt{2} &0\\
0 & 0 &\sqrt{2}\\
0 & 0 &0
\end{pmatrix}\\
\end{equation}
and from \eqref{Wexplicit}
\begin{equation}\label{sl3repW}
\begin{aligned}
&W_0=\begin{pmatrix}
\frac{1}{3} & 0 &0\\
0 & -\frac{2}{3} &0\\
 0& 0 &\frac{1}{3}
\end{pmatrix}\qquad W_1=\begin{pmatrix}
0 & 0 &0\\
 -\frac{1}{\sqrt{2}}& 0 &0\\
 0& \frac{1}{\sqrt{2}} &0
\end{pmatrix}\qquad W_{-1}=\begin{pmatrix}
0 & \frac{1}{\sqrt{2}} &0\\
0 & 0 &-\frac{1}{\sqrt{2}}\\
0 & 0 &0
\end{pmatrix}\\
\noalign{\vskip.2cm}
&\qquad\qquad\qquad W_{2}=\begin{pmatrix}
0 & 0 &0\\
0 & 0 &0\\
 2& 0 &0
\end{pmatrix}\qquad\qquad W_{-2}=\begin{pmatrix}
0 & 0 &2\\
 0& 0 &0\\
 0& 0 &0
\end{pmatrix}\nonumber
\end{aligned}
\end{equation}
The normalization factors (\ref{orthor}) are
\begin{equation}
\begin{aligned}
t^{(2)}_0=2 \qquad  t^{(3)}_{0}=\frac{2}{3} \qquad 
t^{(2)}_1=-4 \qquad t^{(3)}_1=-1 \qquad t^{(3)}_2=4
\end{aligned}
\end{equation}

\section{Some explicit results for $\textrm{SL}(N)$}\label{sec:SL(N)}

In this appendix we collect some explicit results of the constructions presented 
in Chapter \ref{sec:Spin3} which are valid for all $N$.  

We decompose elements of the Cartan subalgebra of 
$\mathfrak{sl}(N)$, such as $\boldsymbol{\Phi}$ into the two bases specified in 
Appendix \ref{sec:conventions}:
\begin{equation}\label{zeromodes}
\boldsymbol{\Phi}\equiv\sum^{N-1}_{i=1} \Phi_i H^{(i)}=\sum^{N}_{s=2}\Psi_{s} W^{(s)}_{0}
\end{equation}
While the closed form expression for the $\textrm{SL}(N)$ group element $M$, 
which implements 
the Miura transformation, is not available, we can
write down the part along the simple root directions
$M=\exp[h_-]=\mathds{1} +h_{-1}+\dots$ where
\begin{equation}
h_{-1}=\sum^{N-1}_{i=1}\frac{1}{\sqrt{k_i}}\, \partial \Phi_i \, E^{(i)}_{+}
=\sum^{N}_{s=2}\frac{1}{s}\, \partial \Psi_s \, W^{(s)}_{-1} 
\end{equation}
Using
\begin{equation}
e^{-\boldsymbol{\phi}}E^{(i)}_{\pm} e^{\boldsymbol{\phi}}=e^{\mp\a_i}E^{(i)}_{\pm} \qquad 
\textrm{with}\qquad\a_i \equiv K_{ij}\phi_j
\end{equation}
the leading terms in the gauge transformed connection become
\begin{equation}
\begin{aligned}\label{aconfslN}
a_{z}&=-\sum_{i}\sqrt{k_i}\,e^{\alpha_i}\, E^{(i)}_{-} 
-\partial \boldsymbol{\bar{\phi}}
+\frac{T}{t^{(2)}_1}\sum_{i} \sqrt{k_i}\,e^{-\alpha_i}\, E^{(i)}_{+} +\dots \\
a_{\bar{z}}&= \bar{\p}\boldsymbol{\phi} 
-\sum_{i}\frac{1}{\sqrt{k_i}}\,\p\bar{\p}\Phi_i\,e^{-\alpha_i}\,E^{(i)}_{+}+\dots
\end{aligned}
\end{equation}
and $\tilde a=-a^\dagger$.
Here $T\equiv q_2(z)+\mathcal{Q}_2(\boldsymbol{\Phi})$, where 
$\mathcal{Q}_{2}$ is the $(zz)$-component of the energy-momentum tensor of 
Toda theory:
\begin{equation}
\mathcal{Q}_2 =-\frac{1}{2}\,\sum_{i,j} K_{ij} \,\partial \Phi_i\, \partial \Phi_j 
+\sum_{i} \partial^2 \Phi_i 
\end{equation}
The terms displayed suffice to compute the leading and subleading terms 
in the $\rho$-expansions of $A$ and $\tilde A$ and the metric-like fields. 
The terms which we have not shown are rather long and we do not have closed 
expressions for all $N$. But for any $N$ they can be worked out straightforwardly, 
following the procedure outlined before.  

Using (\ref{aconfslN}) we can give the leading terms of the 
$\rho$-expansion of the bulk dreibein (cf. \eqref{FGexpansion-e})
\begin{equation}\label{confdreib}
\begin{aligned}
&\ez=-L_0\, d\rho-\frac{1}{2}\sum^{N-1}_{i=1}\sqrt{k}_i\,[e^{\a_i}\,E^{(i)}_{-}\, dz 
+h.c.]\qquad\qquad\eone=0\\
&\etwo=\frac{1}{2}\sum^{N-1}_{i=1}\sqrt{k}_i\,[e^{-\a_i}\, E^{(i)}_{+}\, dt_i 
+h.c.]
\end{aligned}
\end{equation}
where we have defined $dt_i$ 
\begin{equation}
dt_i\equiv \frac{1}{t^{(2)}_1}T dz - \frac{1}{k_i}\partial\bar{\partial}\Phi_i\,d\bar{z}  
\end{equation}
It is now straightforward to compute the bulk metric. 
The first few terms in its FG expansion (\ref{FGexpandG}) are
\begin{equation}\label{gconfsl3}
\begin{aligned}
&\gz=\frac{1}{2 t^{(2)}_0} \, (\sum_{i}k_i \,e^{\mathcal{A}_i}) \, dz d\bar z  
\qquad \qquad \qquad \textrm{with} \qquad \mathcal{A}_i\equiv \alpha_i + \bar{\alpha}_i\\
&\gt=-\frac{1}{2t^{(2)}_0}(\sum_{i} k_i \,dt_i)\,dz+c.c.
=(\partial\bar{\partial}\Phi_L)
\,dz d\bar{z}+\frac{T}{2 t^{(2)}_0}\,dz^2+\frac{\bar{T}}{2t^{(2)}_0}\,d\bar{z}^2\\
&\gf=\frac{1}{2 t^{(2)}_0} \, \sum_{i}k_i \,e^{-\mathcal{A}_i}\,dt_i d\bar{t}_i  
\end{aligned}
\end{equation}
where, using $t^{(2)}_1=-\sum_{i} k_i$,
\begin{equation}
\Phi_L=\frac{1}{t^{(2)}_0}{\rm tr}\big(L_0 \boldsymbol{\Phi}\big)
=\frac{1}{t^{(2)}_0}\sum_{i}\Phi_i
\end{equation}
The 
$i_1 i_2 \dots i_s$ components of the bulk spin-$s$ field has a 
$\rho$-expansion that starts at $\mathcal{O}(\rho^{-s})$/$\mathcal{O}(\rho^0)$ 
for even/odd spin. The component $\Phi^{(s)}_{\rho\dots \rho ij}$ 
component has an $\rho$-expansion
\begin{equation}
\Psi^{(s)}_{\rho\dots \rho ij}
=\frac{1}{\rho^{s}}\sum^{s-1}_{n=0}\rho^{2n}\,\psitwon_{\rho\dots\rho ij}
\end{equation}
Finally we mention that, in analogy to the fact that 
$g^{\scriptscriptstyle(2)}_{z\bar{z}}=\partial \bar{\partial} \Phi_L$ 
is proportional to the Weyl anomaly in conformal gauge (cf. (\ref{gconfsl3})), 
$\psi^{\scriptscriptstyle(2)}_{\rho \dots \rho z\bar{z}}
\sim \partial \bar{\partial} \Phi_{s}$ is proportional to the spin-$s$ 
$\mathcal{W}$-Weyl anomaly
in the same gauge.

\medskip

The construction of the Schwarzian derivatives for $\mathfrak{sl}(N)$ can be easily 
described. One uses the Miura transformation to find the higher-spin 
charges $Q_s$ in terms of the zero modes $\p\Phi_i$, $i=1,\dots,N-1$ along the 
Cartan directions, cf. \eqref{zeromodes}. One then replaces
\begin{equation}
\p\Phi_i=\sum_j(K^{-1})_{ij}{\p^2 e_j\over\p e_j} 
\end{equation}
where $K^{-1}$ is the inverse Cartan-Matrix of $\mathfrak{sl}(N)$ and $e_i$ are 
the functions along the simple root directions which appear in the Gauss
decomposition of the general group element. They were called $e_+$ and $f_+$ in 
\eqref{basis}. From the $\psi_i$, which are a basis of solutions of the 
$N$-th order holomorphic differential equation which generalizes \eqref{eqpsi3z}, 
one forms the $N-1$ ratios $u_i=\psi_i/\psi_N$. The relation between these 
ratios and the $e_i$ is \cite{Bilal:1990wn}
\begin{equation}
e_{N-1}=u_{N-1}\,,\quad
e_{N-2}={1\over\p e_{N-1}}\p u_{N-2}\,,\quad\dots\quad
e_{N-i}={1\over\p e_{N-i+1}}\p{1\over\p e_{N-i+2}}\p\cdots\p{1\over\p e_{N-1}}
\p u_{N-i}
\end{equation} 
Equations \eqref{gl3} and \eqref{uz} generalize to 
\begin{equation}
u_i\to{
\sum_{j=1}^N\,
c_{ij}\,u_j\over
\sum_{j=1}^N\,
c_{Nj}u_j}\,,\qquad i=1,\dots,N-1\,,
\qquad \{c_{ij}\}\in GL(N)
\end{equation}
and 
\begin{equation}
u_i=\frac{
\sum_{j=1}^N\,
c_{ij}\,\frac{z^{N-j}}{(N-j)!}}{
\sum_{j=1}^N\,
c_{Nj}\,\frac{z^{N-j}}{(N-j)!}}
\end{equation}
%
respectively. We have checked these statements explicitly for $SL(4)$. For this case 
we have also found the Beltrami equations, but they are too long to present here, as 
are the explicit expressions for the charges. 

\end{appendix}


\end{document}